\documentclass[aps,pra,superscriptaddress,twocolumn,eps2pdf,notitlepage,10pt]{revtex4-2}

\usepackage[utf8]{inputenc}

\usepackage{hyperref}
\usepackage{graphicx}
\usepackage{amsmath}
\usepackage{amssymb}
\usepackage{dsfont}
\usepackage{cleveref}
\usepackage{braket}
\usepackage{enumerate}
\usepackage[ruled,vlined]{algorithm2e}
\usepackage{algpseudocode}
\usepackage{booktabs}
\usepackage{subfigure}

\crefname{table}{table}{tables}
\Crefname{table}{Table}{Tables}
\crefname{figure}{Fig.}{Figs.}
\Crefname{figure}{Fig.}{Figs.}
\crefname{equation}{Eq.}{Eqs.}
\Crefname{Equation}{Equation}{Equation}

\begin{document}

\title{Reducing quantum error correction overhead using soft information}
\author{Joonas Majaniemi}\email{joonas.majaniemi@riverlane.com}
\affiliation{Riverlane, Cambridge, CB2 3BZ, UK}
\author{Elisha S. Matekole}\email{esmatekole1@gmail.com}
\affiliation{Riverlane, Cambridge, Massachusetts 02142, USA}
\date{March 17, 2026} 

\begin{abstract}
    Imperfect measurements are a prevalent source of error across quantum computing platforms, significantly degrading the logical error rates achievable on current hardware. To mitigate this issue, rich measurement data referred to as soft information has been proposed to efficiently identify and correct measurement errors as they occur. In this work, we model soft information decoding across a variety of physical qubit platforms and decoders and showcase how soft information can make error correction viable at lower code distances and higher physical error rates than is otherwise possible. We simulate the effects of soft information decoding on quantum memories for surface codes and bivariate bicycle codes, and evaluate the error suppression performance of soft decoders against traditional decoders. Our simulations show that soft information decoding on near-term devices can provide up to $11\%$ higher error suppression on superconducting qubits and up to $20\%$ stronger error suppression on neutral atom qubits. These accuracy gains correspond to $13\%$ and $33\%$ reductions in the physical qubit footprint of superconducting and neutral atom devices respectively when operating at a logical error rate of $10^{-6}$, showcasing that soft information is a powerful tool for reducing the cost and complexity of large-scale fault-tolerant quantum computers.
    
\end{abstract}

\maketitle

\section{Introduction}

    Quantum error correction (QEC) provides a pathway to high-fidelity operations on large-scale quantum computers \cite{google2024quantum, bluvstein2024logical, paetznick2024demonstration}, a key requirement for reaching a quantum advantage on noise-limited hardware. Provided error rates in the underlying physical operations are below a certain threshold, QEC enables the encoding of higher-fidelity logical qubits into noisy physical qubits \cite{kitaev2003fault-tolerant}. By increasing the physical qubit overhead, QEC achieves exponential suppression of errors in the logical qubits \cite{aharonov2008fault-tolerant}. The extent of error suppression depends on the error correcting code chosen, the quality of the underlying qubits, and importantly, the accuracy of the classical decoder used for deciphering corrections to the logical state. While recent demonstrations have shown fault-tolerant logical operations and logical error rates below physical qubit error rates \cite{google2024quantum, bluvstein2024logical, paetznick2024demonstration}, near-term error-corrected devices are far away from the logical error rates below $10^{-9}$ that are required for commercially viable applications \cite{beverland2022assessing}. Even with state-of-the art device fidelity, the number of physical qubits needed to encode a commercially viable high-fidelity logical qubit is prohibitively large on current hardware \cite{mohseni2025build}.

    Improvements in classical decoders can offer a shortcut to stronger error suppression, but the accuracy of the decoder is often limited by how much information it has about the underlying system \cite{bausch2024learning}. In standard QEC schemes such as surface code quantum computing \cite{dennis2002topological, kitaev2003fault-tolerant}, successive rounds of stabiliser measurements are performed to track the state of a logical qubit against potential errors. This measurement data is typically converted from an analogue to a digital representation and classified to a binary measurement outcome before being passed to a classical decoder \cite{google2023suppressing}. However, the binary classification process is sub-optimal as it throws away valuable information about the measurement process that can be used to improve decoding accuracy \cite{anjou2021generalised, pattison2021improved}. By using a probability estimator that captures the likelihood of a given measurement signal originating from state $\ket{0}$ or $\ket{1}$, denoted as soft information, the decoder can more accurately track measurement errors over the course of a QEC protocol, enhancing the code threshold and making the decoder more resilient to classification errors.

    Prior simulations have shown that soft information can achieve a higher error-correction threshold on the surface code than traditional protocols \cite{pattison2021improved}, and a number of experimental works have confirmed that the technique yields lower logical error rates in real-world systems \cite{ali2024reducing, caune2024demonstrating, Xue2020repetitive, hanisch2024soft}. The primary question that had been left to answer is whether soft information remains advantageous when moving towards large-scale future systems, and whether the previously shown improvements translate to other qubit platforms and error-correcting codes. It also remained an open question whether soft decoding is compatible with the high-throughput, low-latency decoding workflows needed for future large-scale error-corrected systems \cite{google2024quantum}. Previous soft decoder demonstrations required extensive computation during runtime \cite{pattison2021improved, hanisch2024soft}, which are potentially incompatible with real-time implementations.

    To fill this gap, we incorporate soft decoding capabilities into two promising real-time decoder architectures --- the local clustering decoder (LCD) \cite{ziad2025local} and belief propagation (BP) \cite{shen2020enhanced, borwankar2020low, muller2025improved}, and benchmark their accuracy on extensive simulations of two leading qubit platforms: superconducting transmon qubits and neutral atom qubits. Although superconducting qubits have enjoyed significant successes towards scalable QEC, neutral atoms have emerged as a competitive alternative thanks to their high fidelity, extensive connectivity and the ability to simultaneously manipulate multi-qubit arrays \cite{radnaev2024universal, bluvstein2024logical, xu2024constant-overhead}. On these platforms, we simulate logical performance of our decoders on the surface code as well as the newly developed bivariate bicycle (BB) codes \cite{bravyi2024high-threshold}, and compare the resource overheads needed to reach target error rates with soft and hard decoders. Our findings show that soft information yields consistently higher rates of error suppression than hard decoding across different error-correcting codes and qubit platforms, unlocking low logical error rates at substantially smaller qubit footprints that would otherwise be needed. Our results are of particular relevance to experimentalists, who we hope will leverage soft information to design faster and more hardware-efficient error-corrected quantum devices in the future.

    For our soft-information-augmented decoders, we use LCD \cite{ziad2025local} to decode the surface code simulations, and the BP-based decoder \cite{shen2020enhanced, borwankar2020low} to decode the BB codes. Our choice of decoders is motivated by their prospects in a real-time experiment --- LCD is optimised for high-throughput implementation, and BP variants such as relay \cite{muller2025improved} have been shown to be compatible with the tight time constraints of real-time decoding.  While most small-scale QEC experiments have been conducted with offline decoders where the syndrome information is decoded after the fact, utility-scale quantum computers will require data processing and decoding to happen at the timescale of syndrome extraction --- requiring scalable and responsive decoders. We note that other decoders can be made compatible with soft information, such as minimum-weight perfect matching \cite{higgott2023sparse} which has been shown to deliver improved error rates in real-world experiments \cite{ali2024reducing, caune2024demonstrating, hanisch2024soft}, as well as network decoders which can be trained with soft information, delivering better error rates than their binary counterparts \cite{bausch2024learning, varbanov2025neural}. More general solvers such as Tesseract \cite{beni2025tesseract} may also be used with soft information by modifying the decoder prior, but these lack the scalability needed for a real-time setting. 

    On both platforms, we evaluate the rate of error suppression, represented by the $\Lambda$-factor \cite{kelly2015state}, that can be achieved with soft information decoding and compare it to traditional hard decoding. We then relate improvements in $\Lambda$ to physical qubit footprint estimates given target logical error rates. We see that in simulated quantum memory experiments, soft decoders provide consistently higher $\Lambda$-factors than hard decoders for both superconducting and neutral atom platforms. In noise regimes where measurement classification errors are unlikely compared to data errors, soft decoding offers next to no advantage, but in more realistic settings where measurement classification is a lower-fidelity operation than gates, soft decoding offers between $11\%$- and $20\%$-improvements in $\Lambda$. This corresponds to $13\%$ and $33\%$ reductions in physical qubits needed to reach a logical error rate of $10^{-6}$ for superconducting and neutral atom platforms respectively. Our modelling shows that the scaling advantages of soft decoding are maintained even when measurement times are shorter than ideal---offering faster QEC cycles without a penalty in the error suppression performance compared to hard decoders. For simulated bivariate bicycle codes, we find soft information to reduce logical error probabilities by an order of magnitude, showcasing that the benefits of the technique are applicable across a wide range of QEC codes.

    The paper is structured as follows. A summary of the soft measurement model for superconducting and neutral atom platforms is described in \cref{sec:meas_model}, and our chosen figures of merit are described in \cref{sec:figures_of_merit}. In \cref{sec:soft_matching}, we study the impact of soft decoding for surface code experiments on both superconducting and neutral atom platforms, leveraging soft and hard variants of LCD as our decoder. In \cref{sec:soft_matching_t_measurement}, we simulate error suppression against the readout duration of superconducting and neutral atom platforms, likewise decoding the results with soft and hard LCD. In \cref{sec:soft_bp} we show how soft-BP decoders benefit from soft measurement information, benchmarking the decoders on BB codes on simulated neutral atom and superconducting platforms. Section \ref{sec:discussion} concludes the work and provides pointers on effectively incorporating soft information into the decoding workflows of future QEC experiments.

\section{Method}\label{method}

    \subsection{Soft measurement model}\label{sec:meas_model}

\begin{figure}[h]
    \centering
    \includegraphics[width=0.85\columnwidth]{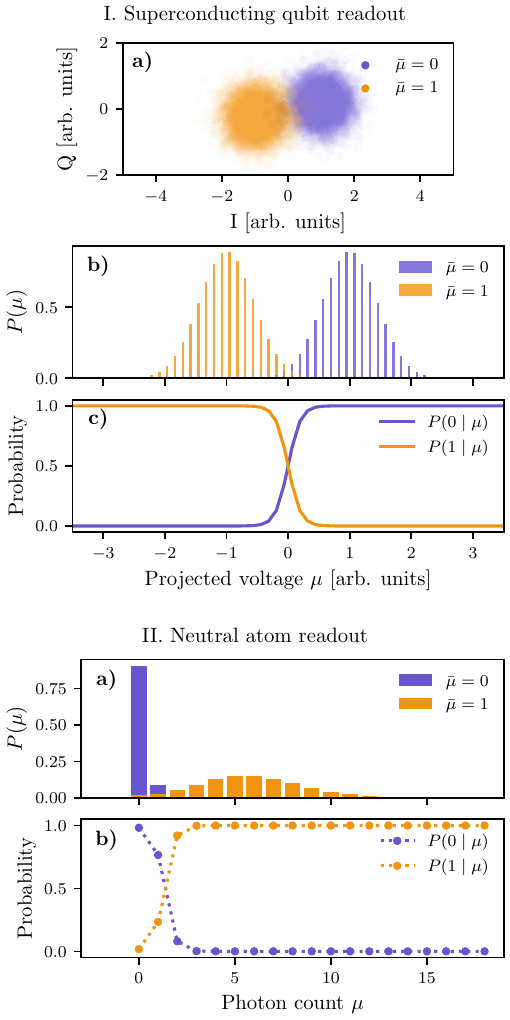}
    \caption{\textbf{Readout model for superconducting and neutral atom qubits.} I. Superconducting readout, a) showing example IQ voltages from a repeated state-preparation and measurement routine for states $\ket{0}$ (in purple) and $\ket{1}$ (in orange), b) the resulting probability density functions when projected along the axis connecting the 0- and 1-centroids, and c) the posterior probability of a measurement outcome $\mu$ given the initial state preparation $\ket{0}$ or $\ket{1}$. II. Neutral atom readout, a) showing photon counts $\mu$ for a bright state $\ket{1}$ (in orange) and a dark state $\ket{0}$ (in purple), and b) the posterior probability of a measurement outcome $\mu$ given either a 0- or 1-state preparation.}
    \label{fig:readout_model}
    \end{figure}

    When measuring the state of a qubit, we treat the resulting signal as a random variable $\mu$ which follows some probability density function $f^{(\bar{\mu}=0)}(\mu)$ for a qubit projected into the $\ket{0}$-state, and a density function $f^{(\bar{\mu}=1)}(\mu)$ for a qubit projected into the $\ket{1}$-state. The type of the measurement distribution depends on the physical mechanism of the measurement process, determined by the qubit type \cite{shea2018fast, krantz2019quantum}. We label the state to which the qubit was projected by the variable $\bar{\mu}$, here referred as the \textit{ideal outcome}, where $\bar{\mu}=0$ and $\bar{\mu}=1$ correspond to states $\ket{0}$ and $\ket{1}$ respectively. To classify a soft measurement response $\mu$ to a binary outcome $\hat{\mu}\in\{0, 1\}$, we calculate from each measurement the posterior probability

    \begin{equation}\label{eq:post_1}
        P(1\mid\mu) = \frac{f^{(\bar{\mu}=1)}(\mu)P(1)}{f^{(\bar{\mu}=0)}(\mu)P(0) + f^{(\bar{\mu}=1)}(\mu)P(1)}
    \end{equation}

    \noindent where $P(0)$ and $P(1)$ are the prior probabilities of the measurement outcomes $\bar{\mu}=0$ and $\bar{\mu}=1$ respectively. In \cref{fig:readout_model}, we show how the measurement signal from superconducting and neutral atom qubits can be mapped to the probability distributions $f^{(\bar{\mu}=0)}(\mu)$ and $f^{(\bar{\mu}=1)}(\mu)$, and how the corresponding posterior probability $P(1\mid\mu)$ and its complement $P(0\mid\mu) = 1 - P(1\mid\mu)$ evolve as a function of the random variable $\mu$.

    The classification outcome for a soft measurement $\mu$, also denoted the \textit{hardened measurement outcome}, is labelled by $\hat{\mu}$. Given a posterior probability $P(1\mid\mu)<1/2$, the classification outcome is $\hat{\mu}=0$, while $P(1\mid\mu) \geq 1/2$ results in a classification outcome $\hat{\mu}=1$. While the physical origin of the measurement probability distribution differs based on the hardware type and can vary qubit-to-qubit, the posterior probability $P(1\mid\mu)$ can be easily mapped to a binary measurement outcome $\hat{\mu}\in\{0, 1\}$ and processed by a decoder without needing to know the explicit form of the measurement response distributions $f^{(\bar{\mu}=0)}$ and $f^{(\bar{\mu}=1)}$. We have opted for un-biased priors $P(0)=P(1)=1/2$ to ensure that our findings are not overly optimistic regarding the effectiveness of soft information: equal priors reflect a case where the decoder is not privy to information regarding the true logical state. To make input to the decoder easy to represent in a high-throughput manner, we express the posterior probability $P(1\mid\mu)$ as an 8-bit unsigned integer. This has been shown to be a sufficient bit depth to retain the full advantages of probabilistic measurement information \cite{hanisch2024soft}. Further optimisation to the soft decoder can potentially be made by incorporating experiment-specific information about the measurement priors $P(0)$ and $P(1)$ into the classification process \cite{bausch2024learning}, but we leave such strategies outside the scope of this work.
    
    In the case of superconducting qubits, shown in \cref{fig:readout_model} I) the measurement signal comes as a dispersive shift of a resonator that is coupled to the qubit, resulting in a unique Gaussian distribution depending on the qubit state \cite{krantz2019quantum}. For neutral atoms, shown in \cref{fig:readout_model} II) we model a non-destructive fluorescence-based state-readout protocol that is compatible with mid-circuit measurements. Measurement is performed using a single photon counting module (SPCM), with the $\ket{0}$- and $\ket{1}$-states exhibiting different amounts of fluorescence \cite{crain2016integrated, noek2013high}. The resulting measurement signal is a unique Poisson distribution of the photon count $\mu$ for the two states \cite{MartinezDorantes2018state-dependent}. We assume for both platforms that readout happens as a perfect quantum non-demolition (QND) measurement, and second-order effects such as measurement back-action on the ancilla qubits or partial measurement of the encoded qubits are omitted from our model. The probability density functions $f^{(\bar{\mu}=0)}(\mu)$ and $f^{(\bar{\mu}=1)}(\mu)$ used for our simulations of superconducting and neutral atom qubits are given in \cref{methods:superconducting_noise_model} and \cref{methods:neutral_atom_noise_model} respectively. In both platforms, there exists a classification boundary at a fixed value of $\mu$ where the measurement outcome is maximally uncertain and classification errors are most likely. On either side of the boundary, measurement outcomes are more likely to be correctly assigned.

    For both platforms, the choice of measurement time  is a key parameter affecting the measurement fidelity---we demonstrate its effect in detail in \cref{sec:soft_matching_t_measurement}. In order to mitigate the random fluctuations inherent in the measurement process, measurement signal is accumulated over a fixed time interval, with long measurement durations resulting in a cleaner signal. However, even with a good choice of measurement time, it is not possible to always distinguish between the two possible states, and cases where an incorrect state is assigned are known as \textit{soft measurement errors} or \textit{soft flips} \cite{pattison2021improved}. Given a soft measurement value $\mu$, the probability of a soft flip is given by
    \begin{equation}
        \label{eq:p-soft}
        {p_\textrm{S}}(\mu) = \min\left[P(0\mid\mu), P(1\mid\mu)\right]
    \end{equation}
    
    \noindent where $P(0\mid\mu) = 1 - P(1\mid\mu)$. The average soft flip probability ${p_\textrm{S}}_{\mid_{\bar{\mu}}}$ for a state projected into the ideal outcome $\bar{\mu}$ is a function of the overlap in the distributions $f^{(\bar{\mu}=0)}(\mu)$ and $f^{(\bar{\mu}=1)}(\mu)$, given by:

    \begin{align}
    \begin{split}\label{eq:soft_overlap}
        {p_\textrm{S}}_{\mid_{\bar{\mu}=0}} &= \int_{f^{(\bar{\mu}=1)} > f^{(\bar{\mu}=0)}} f^{(\bar{\mu}=0)}(\mu) \text{d}\mu \\
        {p_\textrm{S}}_{\mid_{\bar{\mu}=1}} &= \int_{f^{(\bar{\mu}=0)} > f^{(\bar{\mu}=1)}} f^{(\bar{\mu}=1)}(\mu) \text{d}\mu
    \end{split}
    \end{align}

    \noindent where the measurement distributions $f^{(\bar{\mu}=0)}(\mu)$ and $f^{(\bar{\mu}=1)}(\mu)$ depend on the measurement time $\tau_{\textrm{M}}$, among other qubit-specific parameters. To capture the frequency of soft errors in our models, we use the average soft flip probability $p_\textrm{S}\!=\!({p_\textrm{S}}_{\mid_{\bar{\mu}=0}} +  {p_\textrm{S}}_{\mid_{\bar{\mu}=1}})/2$.

    When binary measurements are used, the duration of the measurement operation is typically optimised for maximum measurement fidelity $1 - p_\textrm{S}$ \cite{heinsoo2018rapid, bengtsson2024model-based}, but simulations show that a shorter readout duration can result in a higher logical fidelity when QEC is applied thanks to a reduction in idling errors during measurement \cite{pattison2021improved}. In our model, used in \cref{sec:soft_matching} and \cref{sec:soft_bp}, we use measurement times $\tau_{\textrm{M}}$ and error probabilities according to current state-of-the art devices from Google Quantum AI \cite{google2024quantum} for superconducting devices, and a $Z$-biased noise model along with an appropriate measurement model for neutral atom based devices \cite{Cong2022FTQC, shea2018fast}.
    
    It is crucial to note that soft measurement flips are a distinct error mechanism compared to bit-flip errors that occur on ancilla qubits prior to measurement. If ancilla qubits are not reset between syndrome extraction rounds, the error $e_\textrm{S}$ from a soft measurement flip shows up in two fault locations separated by two time steps \cite{ali2024reducing, geher2025toreset}, whereas a bit-flip error $e_\textrm{B}$ triggers a pair of fault locations separated by a single time step. This difference stems from the fact that a soft measurement error does not affect the true state of the ancilla qubit, while a bit-flip error does. Given a soft measurement flip channel with probability $p_\textrm{S}$ and a measurement bit-flip channel with probability $p_\textrm{B}$, a QEC circuit without resets will result in a decoding graph with two distinct edges $e_\textrm{S}\neq e_\textrm{B}$. In the presence of resets, the two edges have the same endpoints and the measurement error channel can be collapsed into a single probability $p_\textrm{M}=p_\textrm{S}(1 - p_\textrm{B}) + p_\textrm{B}(1 - p_\textrm{S})$. The distinct nature of soft measurement errors compared to other measurement error mechanisms plays an important role in designing optimal syndrome extraction circuits ~\cite{geher2025toreset}.

    \subsection{Figures of merit}\label{sec:figures_of_merit}

    Our metric of interest is the error suppression rate $\Lambda$, which quantifies how much the logical error rate of a given QEC code decreases when the code distance is increased \cite{kelly2015state}. The distance, denoted $d$, defines how many errors are required for the code to experience an undetectable logical failure, with bigger distances making the code more resilient to errors. The logical error rate $\epsilon_d$ per QEC round of a code with distance $d$ is suppressed according to 

    \begin{equation}\label{eq:lambda}
        \epsilon_d \propto \Lambda^{-\frac{d+1}{2}}\quad.
    \end{equation}

    Although experimental works have seen $\Lambda$-values up to $\Lambda \approx 2$ for quantum memories on superconducting devices \cite{google2024quantum}, error suppression factors of $\Lambda > 10$ are frequently used in costing assessments of large-scale quantum computation \cite{beverland2022assessing}. In this work, we model system performance across the fidelity landscape, simulating quantum processing units with performance ranging from the current regime of $\Lambda \approx 1$ to speculative near-future systems with $\Lambda \approx 5$.
    
    To evaluate $\Lambda$, we generate instances of rotated planar codes with varying code distances $d=\{5, 7, 9\}$ and simulate the logical fidelity of a quantum memory experiment taking a fixed number $T=10$ syndrome extraction rounds. By decoding the resulting syndrome measurement data, we compare the state of the logical qubit at the end of the experiment with decoder corrections applied against the known state of the logical qubit. By repeating the experiment over multiple shots, we obtain the logical error probability $P_{\textrm{L}}(T)$ for each code distance $d$ and number of rounds $T$. To extract $\Lambda$ from this data, we first use the relation
    \begin{equation}\label{eq:p_l_per_round}
        P_{\textrm{L}}(T) = \frac{1}{2}\left[1 - (1 - 2\epsilon_d)^T\right]
    \end{equation}
    
    \noindent to compute the logical error rate per round $\epsilon_d$ \cite{google2024quantum}, and then use a linear fit of $\log(\epsilon_d)$ to compute $\Lambda$ --- the details of this process are given in \cref{methods:lambda_fitting}. For consistency, we always use a fitted $\Lambda$ as opposed to a distance-specific value and omit the $d=3$ data point due to finite-size effects \cite{google2021exponential}.

    Given some $p_0$ and $\Lambda$ as obtained from a curve fit, we evaluate physical qubit footprints required for a given target error rate by solving \cref{eq:p_l_per_round} for $d$ given a fixed $P_\textrm{L}$, assuming $T=d$ syndrome extraction rounds \footnote{This assumes lattice surgery as the means of logical computation. For qubit platforms such as neutral atoms that are compatible with transversal logic, only $\mathcal{O}(1)$ rounds are needed for logical operations and one can achieve the same logical error rate with a lower-distance code}. Using the definition of a quantum operation (Quop)~\cite{preskill2025megaquop}, we pick two key logical error rate regimes as benchmarks for device performance:

    \begin{itemize}
        \item \textbf{KiloQuop}: $P_L=10^{-3}$, corresponding to $\sim10^3$ error-free operations. This is an error regime that near-future machines are likely to achieve.
        \item \textbf{MegaQuop}: $P_L=10^{-6}$, corresponding to $\sim10^6$ error-free operations. This error regime is much harder to attain, and sets the benchmark for future devices.
    \end{itemize}

    Given a target regime from the above, we estimate the physical qubit footprint $N$ of a rotated planar surface code for each decoder according to $N=2d_\textrm{min}^2-1$ where $d_\textrm{min}$ is the smallest odd distance that satisfies a logical error rate per $d$ rounds of QEC lower than $10^{-3}$ or $10^{-6}$ for the KiloQuop and MegaQuop regimes, respectively. Our estimates are based on the fitted $\Lambda$, meaning that footprint values that require $d>9$ are extrapolations from the fit.

    In the case of bivariate bicycle codes, we do not extract a $\Lambda$-value, as we would need codes that share the same defining bivariate polynomials. The BB codes reported in Ref.~\cite{bravyi2024high-threshold} were found by a numerical search and do not all belong to the same family. Hence, to study the impact of soft information on these codes we consider the logical error probability $P_{\textrm{L}}$ as the figure of merit.

\section{Results}\label{results}

    \subsection{Soft decoding with LCD}\label{sec:soft_matching}

    \subsubsection{Quantum memory performance}\label{sec:soft_matching_qmem}

    To evaluate decoding accuracy, we simulate quantum memory experiments for superconducting and neutral atom qubits under a circuit-level noise model using the fast Clifford simulator Stim \cite{gidney2021stim}. Our noise models have two parameters---the two-qubit gate depolarising error probability $p$ and the average soft measurement error probability $p_\textrm{S}$. In the case of superconducting qubits, we use a variant of the SI1000 model \cite{gidney2021fault-tolerant} to parametrise the noise channels in our simulations, described in detail in \cref{methods:superconducting_noise_model}. We evaluate logical performance under physical error rate regimes between $p=0.2\%$ and $p=0.7\%$, reflecting recent real-world benchmarks of transmon qubits operating near the error-correction threshold \cite{google2024quantum, piskor2025realquantum}. The threshold (estimated at $p=0.6\%$ in Ref. \cite{pattison2021improved}) provides an upper bound on the effectiveness of soft decoding for the surface code, so we focus our analysis on the sub-threshold scaling behaviour. For our neutral atom simulations we use a $Z$-biased error model with a bias factor of $100$, leading to $Z$-errors with probability $p/3$ and $X$- and $Y$-errors with probability $p/300$. The full noise model for neutral atoms is given in \cref{methods:neutral_atom_noise_model}. Here, we sweep physical error rates from $p=0.5\%$ to $p=1\%$, reflecting the two-qubit gate error rates of recent experiments on neutral atom qubits \cite{Evered2023CZfidelityNA, radnaev2024universal}.
    
    We leverage Stim to generate ideal measurement outcomes $j$ that include all noise channels except for measurement classification errors and pass these to the platform-specific measurement distributions $f^{(\bar{\mu}=j)}(\mu)$ to sample soft measurement values $\mu$. Given the soft measurement samples we evaluate the posterior probability $P(1\mid\mu)$ via \cref{eq:post_1}, which is fed to our decoder along with the syndrome extracted from the hardened measurement outcomes $\hat{\mu}$.
    
    We study error suppression behaviour in two regimes of interest, characterised by the ratio of the soft flip probability $p_\textrm{S}$ and the two-qubit gate depolarising error probability $p$:
    \begin{itemize}
        \item Low-probability soft flips (\textbf{LF}): $p_\textrm{S}=p$
        \item High-probability soft flips (\textbf{HF}): $p_\textrm{S}=5p$
    \end{itemize}
    \noindent In the first regime (LF), we assume that qubit state assignment fidelity can be significantly improved in comparison to other error modes, leading to a low ratio $p_\textrm{S}/p=1$. In the second regime (HF), qubit classification remains a lower-fidelity operation than gates, leading to a ratio of $p_\textrm{S}/p=5$ which is a realistic estimate in line with the metrics seen for current experimental realisations of mid-circuit measurements on superconducting platforms \cite{google2024quantum} and neutral atoms \cite{graham2023NA, nikolov2023RBNA}. In addition to measurement errors caused by soft flips, our circuit-level noise model includes measurement bit-flip errors with probability $p$, leading to an overall measurement error probability of $2p$ and $6p$ for the LF- and HF-regimes respectively at $p\ll1$. We summarize the sub-threshold scaling of soft information decoders and their traditional binary-information counterparts in \Cref{table:performance}.
    
    \begin{figure}[h]
    \centering
    \includegraphics[width=0.85\columnwidth]{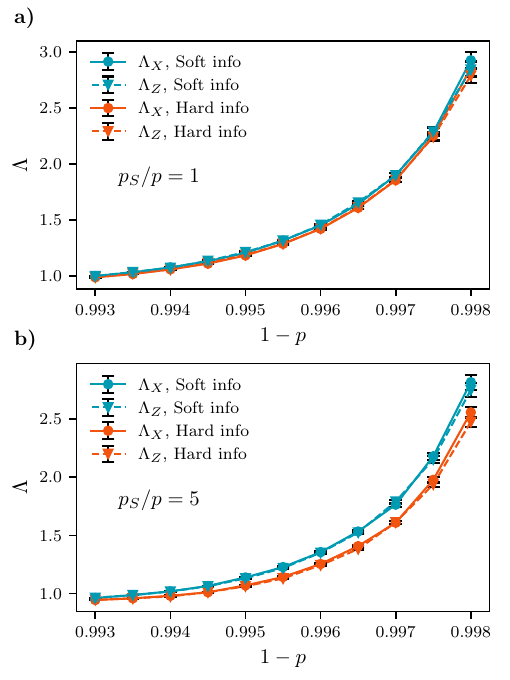}
    \caption{\textbf{LCD on superconducting qubits: error suppression rate $\Lambda$ versus physical qubit fidelity $\mathcal{F}=1 - p$}. Data from a simulated quantum memory experiment with $T=10$ rounds of syndrome extraction and $N=10^6$ shots per data point. In panel a), we show $\Lambda$ for a noise regime where soft flips are sub-dominant $p_\textrm{S}/p = 1$, and in panel b) we show $\Lambda$ when soft flips are a significant component of the error model $p_\textrm{S}/p=5$. Error bars in each plot correspond to hypotheses with a likelihood within a factor of 10 of the maximum likelihood hypothesis, given the sampled data.}
    \label{fig:sc_lambda_vs_fidelity}
    \end{figure}
    
    Starting with superconducting qubits, \cref{fig:sc_lambda_vs_fidelity} shows the effect of the measurement classification error probability $p_\textrm{S}$ and the two-qubit gate fidelity $\mathcal{F}=1 - p$ on the error suppression factor $\Lambda$. For these simulations, we took $10^6$ shots per data point. As qubit fidelity improves, increasing the code distance has a more significant suppressive effect on the logical error rate, leading to larger $\Lambda$-factors with high $\mathcal{F}$. In the LF-regime, shown in \cref{fig:sc_lambda_vs_fidelity} a), the introduction of soft information to the decoder does not significantly affect $\Lambda$ for any of the physical error rates shown. At $p=0.2\%$ we have a minor $(2.3 \pm 0.1)\%$ increase in $\Lambda$ when averaged over the $X$- and $Z$-basis, with the relative gain remaining at $\sim2\%$ all the way up to $p=0.5\%$. This marginal boost in $\Lambda$ does not significantly improve the physical qubit overhead of an error-corrected device, demonstrating that in a setting where soft measurement errors are sub-dominant ($p_\textrm{S}\sim p$), the gain from soft information decoding for superconducting qubits is minimal.
    
    In the HF-regime, plotted in \cref{fig:sc_lambda_vs_fidelity} b), the pattern changes and soft information shows a notable increase in $\Lambda$. At $p=0.2\%$ the $\Lambda$-improvement thanks to soft decoding is $(11 \pm 1)\%$, and remains steady around  $\sim10 \%$ up to $p=0.4\%$. In the near-threshold regime of $p>0.4\%$, the relative impact of soft information on $\Lambda$ decreases---however at these high error rates the difference in logical error rate remains significant. The improved sub-threshold scaling seen in the HF-regime gives a positive indication that soft decoding can be used to reduce the qubit overhead of large-scale fault-tolerant devices. An increase in $\Lambda$ with soft decoding can be used to achieve the same target logical error rate as a hard decoder while using a lower code distance, corresponding to fewer physical qubits needed. The difference in results for the LF- and HF-regimes demonstrates that the soft measurement flip probability $p_\textrm{S}$ is a key parameter that determines whether soft decoding can provide a scaling advantage for superconducting qubits in the long term.

\begin{figure}[htbp]
    \centering
    \includegraphics[width=0.85\columnwidth]{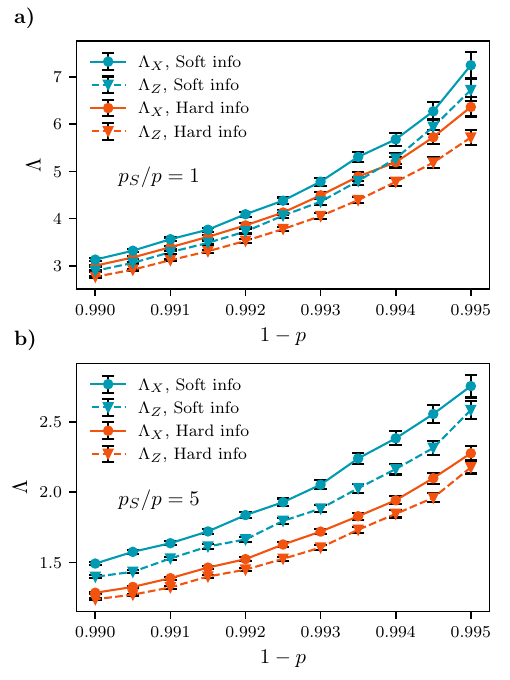}
    \caption{\textbf{LCD on neutral atom qubits: error suppression rate $\Lambda$ versus physical qubit fidelity $\mathcal{F}=1 - p$}. Data from a simulated quantum memory experiment with $T=10$ rounds of syndrome extraction and $N=10^7$ shots per data point in panel a) and $N=10^6$ in panel b). In panel a), we show $\Lambda$ for a noise regime where soft flips are sub-dominant $p_\textrm{S}/p = 1$, and in panel b) we show $\Lambda$ when soft flips are a significant component of the error model $p_\textrm{S}/p=5$. Error bars in each plot correspond to hypotheses with a likelihood within a factor of 10 of the maximum likelihood hypothesis, given the sampled data.}
    \label{fig:na_lambda_vs_fidelity}
    \end{figure}

    In \cref{fig:na_lambda_vs_fidelity} we plot $\Lambda$ against the physical qubit fidelity $\mathcal{F}=1-p$ for a surface code memory simulation of neutral atom qubits, taking $10^7$ shots per data point in the LF-regime and $10^6$ shots per data point in the HF-regime. For simplicity and ease of comparison with the superconducting qubit simulations, we use square surface codes for these examples despite the high noise bias. As expected, improving qubit fidelities on neutral atoms lead to higher $\Lambda$-values. Due to the strong $Z$-bias and the near-lack of idling errors in the neutral atom noise model, the maximum $\Lambda$ in the LF-regime at $p=0.5\%$ is up to $\Lambda = (7.0 \pm 0.4)$ for the soft decoder and $\Lambda = (6.0 \pm 0.3)$ for the hard decoder. We note that extracting high $\Lambda$-values at low error rates is challenging due to the high number of shots required, leading to larger error bars. In the LF-regime, the improvement in $\Lambda$ for soft versus hard decoding ranges from $(16\pm1)\%$ at $p=0.5\%$ to $(4.4\pm0.1)\%$ at $p=1\%$, indicating that the biggest gains in error suppression can only be achieved at high qubit fidelity.

    In the HF-regime, shown in \cref{fig:na_lambda_vs_fidelity} b), we see a consistent relative uplift in $\Lambda$ for soft decoding compared to hard decoding, ranging from $(20\pm1)\%$ at $p=0.5\%$ to $(14.6\pm0.2)\%$ at $p=1\%$. While absolute $\Lambda$ is much lower here than in the LF-regime due to the significant increase in measurement errors, the advantage of soft decoding is clear in this regime. The improvement is consistent across the physical error rate range, showcasing how soft information promises substantially enhanced error suppression rates on noisy near-term neutral atom-based hardware.

    Given the differences in $\Lambda$-factors seen for the soft and hard decoders on both platforms, we evaluate the physical qubit footprints that are required to achieve the KiloQuop and MegaQuop regimes using a rotated planar surface code. Shown in \cref{table:footprint}, we look at the low-noise regimes of $p=0.2\%$ and $p=0.3\%$ for superconducting qubits and $p=0.5\%$ for neutral atom qubits. We observe that in the LF-regime there are only minor differences in qubit count for soft versus hard decoders. However, in the HF-regime we see a clear advantage for soft decoding in terms of physical qubit counts for both superconducting and neutral atom platforms. In the superconducting case, for $p=0.2\%$ a system decoded using soft LCD can reach the KiloQuop regime with $N=241$ physical qubits, as opposed to $N=337$ qubits that are needed with hard information. In the MegaQuop regime, a soft-decoding system needs $1457$ physical qubits, $13\%$ fewer than the $1681$ needed by a hard decoder. For neutral atoms we find a similarly strong reduction in footprint, particularly in the MegaQuop regime, which takes $N=1457$ qubits for a soft decoder in the HF-regime at $p=0.5\%$ compared to a hard decoder needing $N=2177$ physical qubits, a $33\%$ reduction. These improvements demonstrate that soft information can be a powerful tool for reducing the cost and complexity of future large-scale error-corrected quantum devices.

    \begin{table*}[t]
        \centering
        \renewcommand{\arraystretch}{1.2}
        \begin{tabular}{cccccc}
            \toprule
            \textbf{Platform} & $\quad p_{\textrm{S}}\quad$ & $\quad p \quad$ & $\Lambda_{\textrm{hard}}$ & $\Lambda_{\textrm{soft}}$ & $\Lambda$-\textbf{improvement} \\
            \midrule
            NA & $p$ & $0.5\%$ & $(6.0\pm0.3)$  & $(7.0\pm0.4)$ & $(16\pm 1)\%$\\
            NA & $p$ & $1\%$ & $(2.89\pm0.03)$  & $(3.01\pm 0.03)$ & $(4.4\pm 0.1)\%$\\
            NA & $5p$ & $0.5\%$ & $(2.2\pm0.1)$  & $(2.7\pm0.2)$ & $(20\pm 2)\%$ \\
            NA & $5p$ & $1\%$ & $(1.26\pm0.02)$ & $(1.44\pm0.03)$ & $(14.6\pm 0.3)\%$\\
            \midrule
            SC & $p$ & $0.2\%$ & $(2.82\pm0.09)$ & $(2.9\pm0.1)$ & $(2.3\pm0.1)\%$\\
            SC & $p$ & $0.3\%$ & $(1.86\pm0.03)$ & $(1.90\pm0.03)$ & $(2.3\pm0.1)\%$\\
            SC & $p$ & $0.5\%$ & $(1.18\pm0.01)$ & $(1.21\pm0.01)$ & $(2.0\pm0.2)\%$\\
            SC & $5p$ & $0.2\%$ & $(2.52\pm0.07)$ & $(2.78\pm0.09)$ & $(10.6\pm0.4)\%$\\
            SC & $5p$ & $0.3\%$ & $(1.61\pm0.02)$ & $(1.78\pm0.02)$ & $(10.4\pm0.2)\%$\\
            SC & $5p$ & $0.5\%$ & $(1.07\pm0.01)$ & $(1.13\pm0.01)$ & $(6.4\pm0.1)\%$ \\
            \bottomrule
        \end{tabular}
        \caption{\textbf{Performance of soft decoders on simulated quantum memory experiments on superconducting (SC) and neutral atom (NA) platforms.} For neutral atom platforms, we use a $Z$-biased noise model with bias factor $100$. Both the neutral atom and the superconducting simulations use circuit-level noise with error rate $p$, as described in \cref{methods:neutral_atom_noise_model} and \cref{methods:superconducting_noise_model} respectively. We vary the amount of classification error $p_\textrm{S}$ in the system and show the average error suppression factor $\Lambda$ for soft and hard variants of the local clustering decoder. Each experiment uses $T=10$ syndrome extraction rounds and is repeated for $N=10^6$ shots, except the NA simulation at $p_{\textrm{S}}=p$ where $N=10^7$.}
        \label{table:performance}
    \end{table*}

    \begin{table*}[t]
        \centering
        \renewcommand{\arraystretch}{1.2}
        \begin{tabular}{cccccccc}
            \toprule
            \textbf{Platform} & \textbf{Decoder} & $\quad p_{\textrm{S}}\quad$ & $\quad p \quad$ &\quad$N^{\textrm{hard}}_\textrm{KiloQuop}$ & $\quad N^{\textrm{soft}}_\textrm{KiloQuop}$ & $\quad N^{\textrm{hard}}_\textrm{MegaQuop}$ & $\quad N^{\textrm{soft}}_\textrm{MegaQuop}$ \\
            \midrule
            NA & LCD & $p$ & $0.5\%$ & $49~(d_{5})$ & $49~(d_{5})$ & $449~(d_{15})$ & $337~(d_{13})$ \\
            NA & LCD &  $5p$ & $0.5\%$ & $337~(d_{13})$ & $241~(d_{11})$ & $2177~(d_{33})$ & $1457~(d_{27})$\\
            \midrule
            SC & LCD &  $p$ & $0.2\%$ & $241~(d_{11})$ & $241~(d_{11})$ & $1249~(d_{25})$ & $1249~(d_{25})$\\
            SC & LCD &  $p$ & $0.3\%$ & $721~(d_{19})$ & $721~(d_{19})$ & $4049~(d_{45})$ & $3697~(d_{43})$\\
            SC & LCD &  $5p$ & $0.2\%$ & $337~(d_{13})$ & $241~(d_{11})$ & $1681~(d_{29})$ & $1457~(d_{27})$\\
            SC & LCD &  $5p$ & $0.3\%$ & $1457~(d_{27})$ & $881~(d_{21})$ & $6961~(d_{59})$ & $4801~(d_{49})$\\
            \bottomrule
        \end{tabular}
        \caption{\textbf{Physical qubit footprint estimates given target logical error rates, based on quantum memory simulations on superconducting (SC) and neutral atom (NA) platforms.} The number of physical qubits $N$ required for a code with distance $d$ (shown in parentheses) to reach a target number of error-free quantum operations (Quops) is estimated based on the rotated planar surface code. For neutral atom platforms we use a $Z$-biased circuit-level noise model with bias factor $100$ detailed in as described in \cref{methods:neutral_atom_noise_model} and for superconducting qubits we use the circuit-level SI1000 noise model with error rate $p$ as detailed in \cref{methods:superconducting_noise_model}. In each simulation, we vary the amount of classification error $p_\textrm{S}$ and extract syndromes for $T=10$ rounds, repeating the process for $10^6$ shots.}
        \label{table:footprint}
    \end{table*}

    \subsubsection{Measurement time optimisation}\label{sec:soft_matching_t_measurement}

    Mid-circuit measurements are integral to the practical realisation of many QEC codes such as the surface code, but the act of measurement is often among the most error-prone operations in current QEC experiments \cite{google2023suppressing, google2024quantum}. Although it is often possible to improve measurement fidelity by setting longer measurement times, this comes at the cost of increased idling errors and a slower logical clock speed due to the additional time needed for syndrome extraction. To mitigate these challenges, soft information decoding promises improved resilience against measurement errors---making it possible to shorten the measurement time in an experiment without incurring a penalty in the logical fidelity. In this section, we study the impact of the measurement time $\tau_{\textrm{M}}$ on the error suppression rate $\Lambda$ by simulating quantum memory experiments on superconducting and neutral atom qubit platforms.

    For superconducting qubits, we take the SI1000 noise model as our baseline circuit-level noise model and introduce a variation to its error channel probabilities that depends on measurement and gate durations, as described in \cref{methods:superconducting_noise_model}. In our parametrisation, the measurement time $\tau_{\textrm{M}}$ alters the probability of depolarising errors while qubits are idling, being measured or reset. The device operation times are set as $\tau_{\textrm{1q}}=20~\text{ns}$ and $\tau_{\textrm{2q}}=40~\text{ns}$ for one-qubit and two-qubit gates respectively and $\tau_{\textrm{R}}=20~\text{ns}$ for resets. The soft measurement response function is likewise dependent on $\tau_{\textrm{M}}$, with a signal-to-noise ratio (SNR) that is proportional to $\tau_{\textrm{M}}$ for measurement times $\tau_{\textrm{M}}\ll \textrm{T}_1$. Due to amplitude damping with $\textrm{T}_1=100~\mu\textrm{s}$, in line with the current state-of-the art devices \cite{google2024quantum}, qubits experience time-dependent decay both while idling and during measurement \cite{ghosh2012surface}. For neutral atom qubits, the dominant source of error is the two-qubit error rate---with idle noise being insignificant. We use the same $Z$-biased circuit-level noise model (see \cref{methods:neutral_atom_noise_model}) as done previously, but the measurement response function is varied according to $\tau_{\textrm{M}}$.

\begin{figure}[h]
    \centering
    \includegraphics[width=0.85\columnwidth]{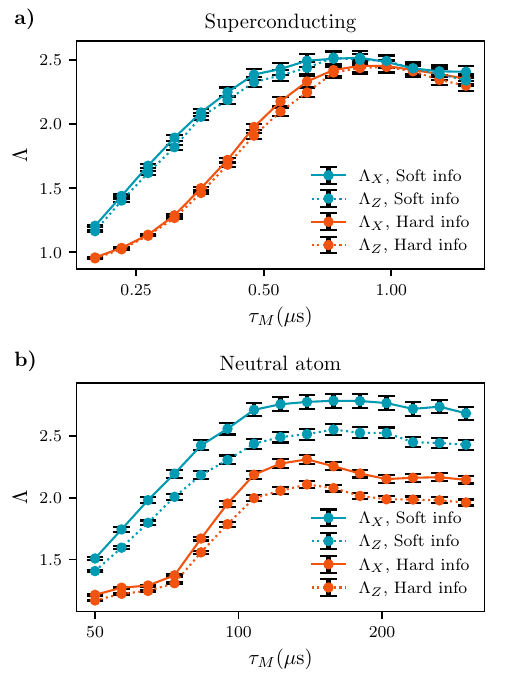}
    \caption{\textbf{LCD: error suppression rate $\Lambda$ versus measurement time $\tau_{\textrm{M}}$}. $\Lambda$ is obtained from simulations of rotated planar code quantum memory experiments with $T=10$ rounds of syndrome extraction. In panel a) we plot $\Lambda$ for superconducting qubits under physical error rate $p=0.3\%$, classification error rate $p_\textrm{S}=5p$ at $\tau_{\textrm{M}}=500~\textrm{ns}$ and $\textrm{T}_1=100~\mu\textrm{s}$. In panel b), we plot $\Lambda$ for neutral atom qubits, with $p=1\%$ and $Z$-bias $100$. The soft flip probability $p_\textrm{S}$ varies based on $\tau_{\textrm{M}}$. On both qubit platforms, a higher $\Lambda$ is reached with soft compared to hard decoding. Error bars in each plot correspond to hypotheses with a likelihood within a factor of 10 of the maximum likelihood hypothesis, given the sampled data.}
    \label{fig:lambda_vs_meas_time}
    \end{figure}

    In \cref{fig:lambda_vs_meas_time}, we vary the code distance $d=\{5, 7, 9\}$ of rotated planar code quantum memories and plot the error suppression rate $\Lambda$ against the measurement time $\tau_{\textrm{M}}$ for superconducting qubits \cref{fig:lambda_vs_meas_time}a) and for neutral atom qubits \cref{fig:lambda_vs_meas_time}b). We vary the superconducting qubit measurement time on the interval $\tau_{\textrm{M}}\in[200~\text{ns},$ $ 1500~\text{ns}]$, and the neutral atom measurement time on the interval $\tau_{\textrm{M}}\in[50~\mu\text{s},$ $ 300~\mu\text{s}]$. Each data point is calculated from $10^6$ shots. On both platforms, we see that for short measurement times the rate of error suppression is decreased. This is explained by the more frequent classification errors that result from a noisier signal. The soft decoders show considerably stronger error suppression than their hard counterparts for small $\tau_{\textrm{M}}$, with average $X$-and $Z$-basis $\Lambda$-factor improved by $(44\pm1)\%$ for superconducting qubits at $\tau_{\textrm{M}}=300~\text{ns}$ and by $(60\pm2) \%$ for neutral atom qubits at $\tau_{\textrm{M}} = 70~\mu\text{s}$.

    As the measurement time $\tau_{\textrm{M}}$ dictates the duration of a QEC cycle, shortening measurements without losing accuracy is a key challenge in the design of error-corrected quantum computers. Taking the maximum $\Lambda$ for the hard decoder as a benchmark, \cref{fig:lambda_vs_meas_time} shows that measurement duration can be shortened by $55 \%$ for superconducting qubits and by $40\%$ for neutral atoms without incurring a loss in error suppression performance. At shorter-than-ideal measurement times, the soft decoder shows much higher $\Lambda$-values than the hard decoder, owing to its better resilience against measurement classification errors. As the superconducting noise model has measurement-time-dependent idling errors, we find that the optimal $\tau_{\textrm{M}}$ for achieving maximum $\Lambda$ across $X$-and $Z$-basis is $35\%$ shorter for the soft decoder than the hard decoder. The difference in $\Lambda$ between the soft and the hard decoder vanishes for long measurement durations, as the improved SNR reduces the frequency of classification errors in the system.
    
    In the neutral atom picture, time-dependent errors during idling are insignificant thanks to the long qubit coherence times and fast single-qubit and two-qubit gates \cite{Evered2023CZfidelityNA, Cong2022FTQC}. When varying the measurement time, only the classification error probability is affected, leading to both the soft and the hard decoders achieving maximum $\Lambda$ at $\tau_{\textrm{M}}\approx150~\mu\textrm{s}$ as seen in \cref{fig:lambda_vs_meas_time} b). Notably, the improvement in $\Lambda$ for the soft decoder compared to the hard decoder is sustained even for $\tau_{\textrm{M}}>200~\mu\textrm{s}$. This is due to soft measurement errors that persist due to bright-to-dark transitions, which are not mitigated by a longer measurement time. In contrast, on superconducting qubits the measurement classification fidelity improves with long measurement times as long as $\tau_{\textrm{M}}\ll\textrm{T}_1$, leading to convergent $\Lambda$-factors for soft and hard decoding when measurement duration is increased.

    \subsection{Soft belief propagation decoding}\label{sec:soft_bp}
    
    In this section, we investigate whether the advantageous scaling behaviour of soft information decoding seen with LCD can be replicated with higher-accuracy decoders. To study this, we take the belief propagation (BP) pre-decoder and make a set of key modifications that allow it to process soft measurement information, explained in detail in \cref{methods: softBP}. While previous works have tackled the problem of soft information in BP decoders in the limited setting of phenomenological noise \cite{raveendran2022soft, berent2024analog}, our methodology is applicable to general circuit-level noise and does not require additional error nodes. We then test these decoders on surface codes and bivariate bicycle codes under a circuit-level noise simulation, comparing the performance of the soft-information-augmented decoders to their hard-information counterparts.

    \subsubsection{Soft information in BP-Matching}
    
    For high-accuracy decoding of surface codes, we choose the belief matching decoder which has been shown to outperform the minimum-weight perfect matching (MWPM) and union find (UF) decoders under circuit-level noise \cite{higgott2023improved}. The BM decoder operates in two stages. First, BP is used to construct a hyper-graph capturing the likely error configurations by estimating the beliefs between the error and check nodes. This is then decomposed into a matching graph which is subsequently decoded by MWPM. Given that the BM decoder contains two stages of decoding, we consider the optimal place to use soft information in the BM decoder. We have two scenarios:

    \begin{enumerate}
        \item Add soft information at the BP-stage to update the hyper-graph and proceed to the decoding step using MWPM---\emph{soft-belief matching}.
        \item Use BP to generate the hyper-graph which is decomposed to a graph and add soft-information at the MWPM stage---\emph{belief soft-matching}.  
    \end{enumerate}

    \noindent Our simulations show that the BM variant with soft information added after the hyper-graph generation performs worse than a BM decoder in which soft information is added to at the matching stage. We postulate that if soft information is not included from the beginning, BP may converge to an inaccurate estimate of posterior probabilities, the damage from which cannot be fully recovered by the soft updates to MWPM. We therefore use the variant with soft information introduced at the belief stage---the soft-belief matching decoder.

\begin{figure}[h]
        \centering
        \includegraphics[width=0.85\columnwidth]{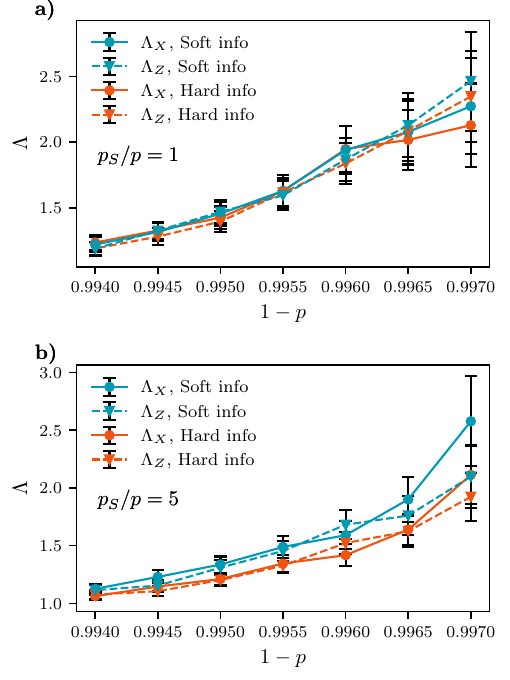}
        \caption{\textbf{BM on superconducting qubits: error suppression rate $\Lambda$ versus physical qubit fidelity $\mathcal{F}=1 - p$}. Data from a simulated quantum memory experiment with $T=10$ rounds of syndrome extraction. Sample size is limited to $N=5\times10^5$ due to slow decoder throughput, causing significant uncertainty in $\Lambda$ at low error rates. In panel a), we show $\Lambda$ for a noise regime where soft flips are sub-dominant $p_\textrm{S}/p = 1$, and in panel b) we show $\Lambda$ when soft flips are a significant component of the error model $p_\textrm{S}/p=5$. Error bars in each plot correspond to hypotheses with a likelihood within a factor of 10 of the maximum likelihood hypothesis, given the sampled data.
        }
        \label{fig:bm_lambda_fidelity}
    \end{figure}

    In \cref{fig:bm_lambda_fidelity}, we evaluate the error suppression performance of a soft-BM decoder on a simulated quantum memory experiment on a superconducting platform. We generate instances of rotated planar codes with code distances $d=\{5, 7, 9\}$, similar to the experiment shown in \cref{fig:sc_lambda_vs_fidelity}, but this time decoded with the soft-BM decoder. The resulting $\Lambda$-factors are recorded in \cref{table:BM_performance}.

    In the LF-regime of \cref{fig:bm_lambda_fidelity}~a), we see statistically insignificant $(6\pm 2)\%$- and $(4\pm 0.5)\%$-improvements in $\Lambda$ when $p=0.3\%$ and $p=0.5\%$ respectively. These results were obtained with $5\times 10^5$ shots for each data point, resulting in relatively large error bars. In the HF-regime, shown in \cref{fig:bm_lambda_fidelity}~b), we see a pattern emerging, with an improvement in $\Lambda$ of to $(16\pm 4)\%$ when $p=0.3\%$ and to $(10\pm1)\%$ when $p=0.5\%$. Again, we observe that the logical error suppression obtained from soft decoding is more robust to increased classification errors than hard decoding. Compared to the results seen with LCD in \cref{sec:soft_matching_qmem}, soft BM achieves higher $\Lambda$-values at comparable error rates---reaching $\Lambda=(2.4\pm0.5)$ at $p=0.3\%$ in the HF-regime versus $\Lambda=(1.78\pm0.03)$ for soft LCD. Similarly to LCD, soft information does not provide an advantage to BM in the LF-regime where $p_\textrm{S}\sim p$, but the performance uplift is significant when soft measurement errors dominate.
    
    The low numbers of samples in this simulation are due to the large decoding time overheads that limit the number of shots taken. There are two main factors that contribute to this. First, the detector error model needs to be updated for each shot with the corresponding soft information. Using circuit-level noise makes the BP part of the algorithm computationally intensive because unlike the matching graph, the circuit-level Tanner graph has a larger number of edges as each qubit might be connected to multiple syndrome nodes. Second, the Ordered Statistics Decoding (OSD) post-processing step used to decode the BP result is highly time-intensive. Overall, even though soft decoding shows a promising uplift in $\Lambda$-factors when $p_\textrm{S}=5p$, the slow decoding speed of BP-Matching makes the technique poorly suited to a high-throughput setting such as a superconducting qubit platform.

    \begin{table*}[t]
    \centering
    \renewcommand{\arraystretch}{1.2}
    \begin{tabular}{cccccc}
        \toprule
        \textbf{Platform} & $\quad p_{\textrm{S}}\quad$ & $\quad p \quad$ & $\Lambda_{\textrm{hard}}$ & $\Lambda_{\textrm{soft}}$ & $\Lambda$-\textbf{improvement} \\
        \midrule
       
        SC & $p$ & $0.3\%$  & $(2.3\pm0.5)$ & $(2.4\pm0.5)$ & $(6\pm2)\%$ \\
        SC & $p$ & $0.5\%$  & $(1.41\pm0.12)$ & $(1.46\pm0.13)$ & $(4\pm0.5)\%$ \\
        SC & $5p$ & $0.3\%$  & $(2\pm0.3)$ &$(2.4\pm0.5)$ & $(16\pm4)\%$ \\
        SC & $5p$ & $0.5\%$  & $(1.21\pm0.08)$ &$(1.32\pm0.1)$ & $(10\pm1)\%$ \\
        \bottomrule
    \end{tabular}
    \caption{\textbf{Performance of soft-BM on simulated quantum memory experiments on a superconducting (SC) platform.} We add soft information into BM decoders, and vary the amount of classification error $p_{\textrm{S}}$, and compare the average logical error suppression factor $\Lambda$ for soft and hard variants of BM. To extract $\Lambda$, we generate instances of rotated planar codes with code distances $d=\{5, 7, 9\}$ and simulate the logical fidelity of a quantum memory experiment taking $T=10$ syndrome extraction rounds. We perform this simulation with $N=5\times 10^{5}$ shots per data point---uncertainties in the resulting values correspond to hypotheses with a likelihood within a factor of $10$ of the maximum likelihood hypothesis, given the sampled data.}
    \label{table:BM_performance}
    \end{table*}

    \subsubsection{Soft information in BP+OSD}

    Bivariate bicycle (BB) codes are quantum low-density parity check (LDPC) codes that exploit long-range qubit connectivity to achieve higher encoding rates than comparably sized surface codes, while maintaining a high error threshold \cite{bravyi2024high-threshold}. This high connectivity needed for their implementation can be easily realised via qubit shuttling on atomic qubits \cite{Pecorari2025}, making neutral atoms a viable candidate for the realisation of these codes in the near term. Alternatively, the degree-6 Tanner graph of the BB codes can be decomposed into two planar subgraphs facilitating their practical implementation on a superconducting qubit architecture.
    
    The Gross code $[[n, k, d]]\!\!=\!\![[144,12,12]]]$ is among the most promising for near-term demonstrations thanks to its high encoding rate. In this section, we analyse the performance of BB codes under a circuit-level noise model for neutral atom qubits as well as the superconducting qubits, decoding the resulting soft measurements with a BP decoder augmented with soft information. To break the degeneracy problem otherwise inherent in the BP algorithm, we combine soft-BP with Ordered Statistics Decoding (OSD) \cite{roffe2020decoding, panteleev2021degeneratequantum, higgott2023improved} as a post-processing step. We note that significantly faster decoders have been proposed \cite{wolanski2024ambiguity, hillmann2024localized, muller2025improved}, however these were not available to us when these simulations were conducted.

    To evaluate error suppression performance, we take the syndrome extraction circuits in \cite{gong2024lowlatencyiterativedecodingqldpc} for two instances of BB codes $[[n, k, d]]\!\!=\!\![[72, 12, 6]]$ and $[[144, 12, 12]]$, and replace the noise in these circuits with an architecture-specific circuit-level noise model. For the neutral atom simulations, we use a $Z$-biased noise model as described in \cref{methods:neutral_atom_noise_model}, while the superconducting simulations have a SI1000 noise model applied, as detailed in \cref{methods:superconducting_noise_model}. Per standard practice we run syndrome extraction circuits for $N_{\textrm{r}}=d$ rounds, and repeat each simulation for a total of $5\times10^5$ shots per data point. In these simulations, we assume qubit connectivity such that all two-qubit gates specified in the syndrome extraction circuits can be realised natively. As before, we evaluate decoder performance under two noise regimes---LF with $p_\textrm{S}=p$ and HF with $p_\textrm{S}=5p$, decoding with both a soft-information-aware BP+OSD decoder and a traditional hard BP+OSD decoder. We set up BP with $100$ BP iterations, \textit{min-sum} as the BP method of choice, combination sweep as the OSD method and an OSD order of $0$. In this work we use parallel BP schedule. For more details on the methods used within BP we refer the reader to Ref.~\cite{roffe2020decoding}.
    
    The measurement error probabilities obtained from soft information are used to dynamically update the error channel probabilities of the BP+OSD decoder using the $\textit{update channel probabilities}$ function in \cite{roffe2022LDPC}. For our figure of merit, we choose the logical error probability $P_{\textrm{L}}$ instead of $\Lambda$ which was used for our surface code simulations. This choice is due to our simulations only involving two BB codes of the same family with distances $d=6$ and $d=12$, which is insufficient for extracting a $\Lambda$-like error suppression metric.
    
\begin{figure*}[ht!]
    \centering
    \begin{minipage}[b]{0.45\textwidth}
    \includegraphics[width=\linewidth]{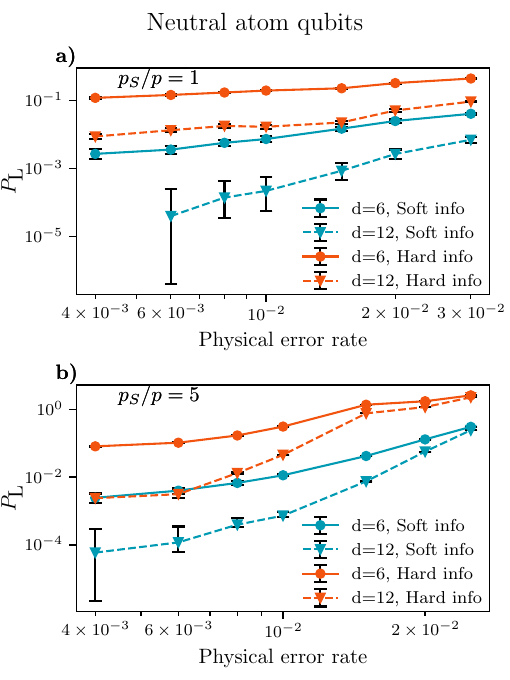}
    \end{minipage}
    \vspace{1em}
    \begin{minipage}[b]{0.45\textwidth}
        \includegraphics[width=\linewidth]{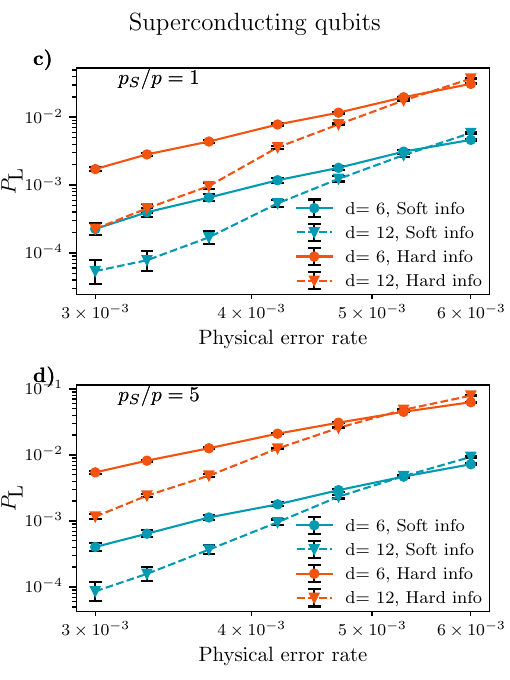}
    \end{minipage}
    \caption{\textbf{Logical error probability $P_{\textrm{L}}$ as a function of the physical error rate $p$ for bivariate bicycle (BB) codes decoded with BP+OSD}. We consider two instances of BB codes: $d\!\!=\!\!6$ for the $[[72,12,6]]$ code and $d\!\!=\!\!12$ for the $[[144,12,12]]$ Gross code. In panels a) and b) we plot the logical error probability $P_{\textrm{L}}$ for simulated neutral atom qubits under two different regimes of soft measurement error probability: $p_\textrm{S}/p=1$ for sub-dominant soft measurement errors, and $p_\textrm{S}/p=5$ for dominant soft measurement errors. Each data point in panels a) and b) uses $5\times 10^{4}$ shots. In panels c) and d) we plot the logical error probability $P_{\textrm{L}}$ for simulated superconducting qubits using $5\times 10^{5}$ shots per data point, with $p_\textrm{S}/p=1$ and $p_\textrm{S}/p=5$ in panels c) and d) respectively. Error bars in all the plots correspond to hypotheses with a likelihood within a factor of 10 of the maximum likelihood hypothesis, given the sampled data.
    }
    \label{fig:bposd_BB}
    \end{figure*}

    In \cref{fig:bposd_BB}, we evaluate decoding accuracy for BP+OSD under the LF and HF soft-measurement flip regimes for both superconducting and neutral atom qubit platforms. We find that in the LF-regime, there is an order-of-magnitude improvement in the logical error rate for superconducting qubits at $p=0.6\%$, and the relative gap between the soft and hard decoders remains consistent all the way down to $p=0.3\%$ for both the distance-$6$ and distance-$12$ codes. In the neutral atom case with heavily $Z$-biased noise, soft decoding decreases the logical error rate by $\sim12\times$ at $p=3\%$ for both the distance-$6$ and distance-$12$ codes. This improvement widens as error rates decrease, and the gap is also distance-dependent: at $p=1\%$ we see a $30\times$ reduction in $p_\textrm{L}$ for $d=6$, and an $80\times$ reduction in $p_\textrm{L}$ for $d=12$. Our results clearly show that soft information in BP+OSD significantly enhances the error suppression ability of BB codes, especially at the high physical error rates likely to be seen in near-term devices.

    \subsection{Experimental considerations}

    \subsubsection{Real-time decoding}

    Our simulations suggest that soft information can reduce the measurement times $\tau_\textrm{M}$ needed for syndrome extraction while simultaneously improving the logical fidelity. To realise this benefit in practice, one needs a classical decoder that is able to cope with the increased rate of syndrome information and can perform soft-information-based graph re-weighting in real time. The long measurement times of neutral atom platforms make decoding speed a secondary concern, but in the superconducting qubit setting the very fast QEC cycle times of $\sim1~\mu \textrm{s}$ dictate very strict timing constraints for the classical decoder. While this challenge may seem daunting, there are several optimisations that can be made to the soft decoding workflow originally proposed in \cite{pattison2021improved} that make it potentially feasible in a real-time application, listed below.
    
    First, the soft measurement probability $P(1\mid\mu)$ can be handled with a fixed precision of $N=8$ bits without compromising logical accuracy, as was found by \cite{hanisch2024soft}. This bandwidth can be further reduced with the use of measurement compression schemes, where measurements whose probability substantially deviates from the high-confidence regions of $0$ or $1$ are allocated $N$ bits, while measurements with a high-likelihood outcome are transmitted as a single bit. This greatly reduces the bandwidth overhead compared to a naive $N$-bit approach. Second, instead of computing the graph modifications that result from the soft measurement samples in a given shot, one can leverage the finite number of possible soft measurement values to pre-compute an edge weight lookup table (LUT). The LUT consists of a map going from a measurement error probability to the new edge weight that needs to be used for the given measurement error mechanism in the decoding graph. Thanks to finite precision, there are only $2^N=256$ possible soft measurement error probabilities that the control system needs to transmit, and as there is a $1$-to-$1$ mapping between a measurement error and a graph refinement event, the entire set of graph modifications only needs to be computed once. Storing the soft-measurement graph refinements in a LUT therefore removes the need for complex arithmetic operations during runtime and paves the way for low-latency soft decoding implementations in dedicated hardware.
    
    In the case of decoding bivariate bicycle codes, recent developments in fast BP-based decoders such as Ref.~\cite{muller2025improved} offer hope that real-time soft information decoding of high-encoding rate qLDPC codes may become possible in the near future. Here too, tracking the soft measurements with finite precision and pre-computing all the necessary hyper-edge modifications on the Tanner graph subject to soft measurement errors may prove essential for a high-throughput implementation.

    \subsubsection{Measurement model fine-tuning}

    When used in practice, it is essential that the soft information model closely matches the real-world measurement distribution, as any deviation will lead to less accurate classification. In a superconducting platform, the measurement signal may deviate from our reference functions shown in \ref{methods:superconducting_noise_model} in a number of ways. Firstly, the presence of leakage may require a three-state classifier which captures the probability of a measurement having originated from either the computational subspace or the leaked $\ket{2}$-state \cite{bausch2024learning, hanisch2024soft}. To mitigate this, leakage-aware decoding schemes \cite{ziad2025local} should be used in tandem with soft information, making the decoder resilient to both measurement classification errors and leakage errors simultaneously. Secondly, due to the non-uniformity of transmon qubits, one needs to fit a unique measurement PDF for every readout in order to achieve accurate classification \cite{ali2024reducing, caune2024demonstrating}. Furthermore, due to the presence of drift, repeated calibration runs of the device are required to ensure that the fitted PDF tracks the real measurement distribution at all times.
    
    On neutral atom platforms, the properties of the imaging pipeline may cause deviations between our model (cf. [Section VI.A.2]) and a real-world measurement response. Optical effects such as vignetting of the image near the edges of the array, brightness crosstalk from neighbouring qubits, or other lens aberrations can decrease the signal-to-noise ratio. Additionally, sensor-based noise sources such as dark current non-uniformity and thermal fluctuations can lead to increased background noise [Radnaev et al. (2024)]. These effects can be mitigated by fitting unique per-qubit probability density functions based on a reference set of measurement values, with each PDFs incorporating optical and sensor-based noise components.
    
\section{Discussion}\label{sec:discussion}

    Our simulations show that soft decoding increases $\Lambda$ compared to hard decoding, making it possible to accomplish the same logical error rates as a hard decoder while using fewer physical qubits. When decoding surface code quantum memory experiments with realistic physical error rates and a high amount of soft measurement flips, soft LCD improves $\Lambda$ by $(20\pm2)\%$ for neutral atoms and by $(11\pm1)\%$ for superconducting qubits. These correspond to $33\%$ and $13\%$ reductions in the physical qubit footprints required for MegaQuop-scale computation on the two platforms respectively---demonstrating that soft information can be a powerful tool for reducing the hardware overheads of error-corrected quantum computation. In addition to the physical space reductions, we also find benefits in soft information for making QEC circuits faster. Shortening measurement times can increase $\Lambda$ for superconducting qubits when decoding with soft information, showing that faster-than-usual QEC circuits can be compatible with good sub-threshold scaling. Furthermore, our simulations employing soft-BM and soft-BP+OSD confirm that the advantageous scaling properties of soft decoding as seen in \cite{pattison2021improved, hanisch2024soft} can be replicated in different QEC codes and decoders.
    
    The extent of the improvement seen with soft decoding across the different platforms and decoders depends on how prevalent soft measurement errors are in the system. Therefore, future improvements in the measurement classification fidelity determine whether soft decoding retains an advantage. Notably, if the mid-circuit measurement classification fidelity is significantly improved over the current state-of-the art, the benefit offered by soft information becomes less dramatic, making the additional engineering cost less justified. However, if reaching low measurement error rates proves to be unfeasible, soft decoding offers a viable compromise for hardware manufacturers, allowing them to trade measurement fidelity for logical fidelity by offloading the additional measurement error processing to the soft decoder. Furthermore, by reducing measurement times through the use of soft information, the duration of syndrome extraction rounds in QEC experiments can be decreased, simultaneously improving the logical fidelity and enhancing the logical clock speed of the error-corrected device.

\section{Conclusion}
    
    While our simulations show that soft information offers significant advantages in the error suppression performance of QEC protocols, key aspects of our results are yet to be verified experimentally. It has been shown for superconducting qubits that soft decoding can achieve a higher $\Lambda$-factor than traditional hard decoding in a repetition code \cite{hanisch2024soft}, but these results include a leakage-decoding component which may obscure the effects of soft information on its own. Larger-scale experiments with real-time decoding such as seen in Ref.~\cite{google2024quantum} could be designed with a fast soft information decoder, offering critical insights into the scalability of soft decoding protocols for future large-scale error-corrected devices. Such work could also quantify the impact of soft decoding on decoding time, likely to be realised with a hardware decoding platform such as those proposed in Refs.~\cite{ziad2025local, liyanage2023scalable}. Experimentalists could also leverage soft information as part of the measurement calibration workflow, confirming if it can be used to tune an error-corrected device for the highest possible logical fidelity. A potentially rich area for further optimisation is the use of high-dimensional classification protocols such as neural networks \cite{vora2024ml}, which may benefit from additional rich input such as the measurement time traces of a superconducting processor. Such advanced classifiers could further improve the fidelity of the measurement posteriors, enhancing decoding accuracy.

    Further experiments need not be constrained to surface codes or repetition codes thanks to the introduction of bivariate bicycle codes \cite{bravyi2024high-threshold}. Experimental realisations of soft decoding on atomic qubits with qLDPC codes present a major milestone in this area, and neutral atoms may be particularly well suited to near-term demonstrations of soft decoding on BB codes thanks to their high qubit connectivity. As prior implementations of soft decoding have been solely focused on solid-state qubits, employing more efficient QEC codes on atomic qubits would confirm the wide applicability of soft decoding techniques for different qubit platforms. The slower operating speeds of atoms likely make any additional time overhead resulting from soft information processing more easily manageable, enabling such QEC demonstrations to be performed in real time. It would also be of great interest to see how an $XZZX$ surface code \cite{bonilla2021xzzx} tailored to the strongly $Z$-biased noise of neutral atoms would perform against a typical square surface code, and whether soft information could be used to further augment performance in this setting.

    In addition to experimental realisations of the protocols studied in this paper, further work is needed to determine how soft information about leakage events, i.e. probabilistic leakage flags determined during the classification process, can be used to decode QEC experiments on superconducting platforms. While leakage can be treated as a discrete event as done in Refs.~\cite{hanisch2024soft, ali2024reducing, ziad2025local}, better logical fidelities may be reached by quantifying the leakage probability as done with the neural network decoder of Ref.~\cite{bausch2024learning}. As graph-based decoders are likely to be required for fast-throughput and low-latency QEC experiments, it remains an open question how to incorporate probabilistic leakage updates into a graph decoding workflow.

\section{Appendix}\label{sec:appendix}

 \subsection{Noise models and simulation parameters}

    Our numerical simulations of QEC circuits leverage the fast Clifford simulator Stim \cite{gidney2021stim} to generate measurement samples. For each qubit platform, we add circuit-level noise according to a physically informed model as detailed in \cref{methods:neutral_atom_noise_model} for neutral atoms and \cref{methods:superconducting_noise_model} for superconducting devices. We treat the Stim-generated measurement samples as the initial outcomes $\bar{\mu}\in \{0, 1\}$ and sample soft measurements $\mu$ from the platform-specific measurement response distributions $f^{(\bar{\mu}=0)}(\mu)$ and $f^{(\bar{\mu}=1)}(\mu)$. All decoding is then performed on the posterior probabilities $P(1\mid\mu)$ for the soft decoders or the hardened outcomes $\hat{\mu}$ for the hard decoders.

    \subsubsection{Superconducting qubit noise model and simulations}\label{methods:superconducting_noise_model}

    We use the SI1000 noise model from Google Quantum AI \cite{google2023suppressing, gidney2021fault-tolerant} as our baseline circuit-level noise model for superconducting qubits. This model is parametrised by the two-qubit gate depolarising error probability $p$, with single-qubit Clifford gates having an error probability $p/10$, initialisation errors in the $Z$-basis having probability $2p$ and $Z$-basis measurement errors having probability $p_\textrm{M}=5p$. In our model, we decompose the measurement error probability $p_\textrm{M}$ into a contribution $p_\textrm{B}$ from a quantum bit-flip error occurring during the measurement process and a contribution with error probability $p_\textrm{S}$ from a soft measurement error at the classification stage. We fix $p_\textrm{B}=p$ and choose $p_\textrm{S}$ based on the noise regime such that the overall measurement error probability satisfies

    \begin{equation}\label{eq:meas_error_decomposition}
        p_\textrm{M} = p\left(1 - p_\textrm{S}\right) + p_\textrm{S}\left(1 - p\right)
    \end{equation}

    \noindent where in the LF-regime, $p_\textrm{S}=p$ leads to $p_\textrm{M}\sim2p$ and in the HF-regime $p_\textrm{S}=5p$ leads to $p_\textrm{M}\sim6p$, both taken in the limit $p\ll1$.

    The average classification error probability $p_\textrm{S}$ is computed from the overlap integral of the soft measurement response functions $f^{(0)}(\mu)$ and $f^{(1)}(\mu)$. Here we define $f^{(0)}(\mu)$ and $f^{(1)}(\mu)$ according to the dispersive readout model described in Refs.~\cite{pattison2021improved, bausch2024learning}. Given a qubit measurement with measurement time $\tau_{\textrm{M}}$, amplitude damping time $\text{T}_1$ and signal-to-noise ratio $\text{SNR}$, the probability density function $f^{(0)}(\mu)$ of the soft outcome $\mu$ for a qubit measured in the ideal outcome $\bar{\mu}=0$ is given by
    
    \begin{equation}\label{eq:f_0}
        f^{(\bar{\mu}=0)}(\mu) = \sqrt{\frac{\text{SNR}}{4\pi}}\exp\left(-\frac{\text{SNR}}{4}\left(\mu - 1\right)^2\right)
    \end{equation}

    \noindent and the probability density function $f^{(1)}(\mu)$ of the soft outcome $\mu$ for a qubit measured in the ideal outcome $\bar{\mu}=1$ is defined as
    \begin{align}\label{eq:f_1}
    \begin{split}
    f^{(\bar{\mu}=1)}(\mu) &= \sqrt{\frac{\text{SNR}}{4\pi}}\exp\left(-\frac{\text{SNR}}{4}\left(\mu + 1\right)^2 - \frac{\tau_{\textrm{M}}}{\text{T}_1}\right)\\
    &- \left[g^{(\bar{\mu}=0)}(\mu) - g^{(\bar{\mu}=1)}(\mu)\right]\frac{\tau_{\textrm{M}}}{4\text{T}_1}\\&\times\exp \left( \frac{1}{4\times\text{SNR}}\left(\frac{\tau_{\textrm{M}}}{\text{T}_1}\right)^2 + \frac{\tau_{\textrm{M}}}{2\text{T}_1}\left(\mu - 1\right)\right)
    \end{split}
    \end{align}
    
    \noindent where $g^{(\bar{\mu}=0)}(\mu)$ and $g^{(\bar{\mu}=1)}(\mu)$ are given by
    
    \begin{align}
        g^{(\bar{\mu}=0)}(\mu) &= \text{erfc}\left[\sqrt{\frac{1}{4 \times \text{SNR}}} \left(\frac{\tau_{\textrm{M}}}{\text{T}_1}\right) + \sqrt{\frac{\text{SNR}}{4}}\left(\mu - 1\right)
        \right]\\
        g^{(\bar{\mu}=1)}(\mu) &= \text{erfc}\left[\sqrt{\frac{1}{4 \times \text{SNR}}} \left(\frac{\tau_{\textrm{M}}}{\text{T}_1}\right) + \sqrt{\frac{\text{SNR}}{4}}\left(\mu + 1\right)
        \right]
    \end{align}
    
    \noindent where $\textrm{erfc}$ is the complementary error function, used to maintain numeric precision when evaluating the probability density function $f^{(1)}(\mu)$ across a wide range of soft outcomes $\mu$. The measurement time relates to the signal-to-noise ratio via 

    \begin{equation}\label{eq:snr}
        \text{SNR} = \frac{\tau_{\textrm{M}}}{2\tau_\textrm{F}}
    \end{equation}
    
    \noindent where $\tau_\textrm{F}$ is a characteristic fluctuation time of the readout signal. In the limit where the measurement time is much shorter than the amplitude damping time $\tau_{\textrm{M}} \ll \text{T}_1$, the classification error probability $p_\textrm{S}$ can be easily calculated from
    
    \begin{equation}
        p_\textrm{S} = \frac{1}{2}\text{erfc}\left(\frac{\sqrt{\text{SNR}}}{2}\right) \quad .
    \end{equation}

    To define our experiments in \cref{sec:soft_matching_qmem}, we fix the two-qubit depolarising error rate $p$ and the soft measurement error probability $p_\textrm{S}$ to our desired values. From these we derive the gate error rates as specified by SI1000, and define the measurement response functions $f^{(\bar{\mu}=0)}(\mu)$ and $f^{(\bar{\mu}=1)}(\mu)$ with a $\text{SNR}$ that satisfies \cref{eq:snr}.

    In the time-dependence analysis of \cref{sec:soft_matching_t_measurement}, we use the SI1000 model as a starting point for our circuit-level noise model and incorporate the measurement time $\tau_{\textrm{M}}$ into the noise channel probabilities as follows. We again start by fixing $p$ and maintain the same error channel probabilities for single-qubit gates, two-qubit gates and reset operations as SI1000. Time-dependence is introduced by incorporating idling noise to qubits that are not acted upon by gates, measurements or resets, with the probability of the idling error channel determined by the specific duration of each operation. We set $\tau_\textrm{1q}=20~\text{ns}$ for single-qubit gates, $\tau_\textrm{2q}=40~\text{ns}$ for two-qubit gates and $\tau_\textrm{R}=40~\text{ns}$ for resets. Based on the duration $\tau$ of each operation, single-qubit Pauli noise with components $(P_X, P_Y, P_Z)$ is added to the idling qubits according to the following probabilities, as taken from Ref.~\cite{ghosh2012surface}:

    \begin{align}
    \begin{split}
        P_X &= \frac{1}{4}\left[1 - \exp\left(-\tau/\textrm{T}_1\right)\right]\\
        P_Y &= \frac{1}{4}\left[1 - \exp\left(-\tau/\textrm{T}_1\right)\right]\\
        P_Z &= \frac{1}{2}\left[1 - \exp\left(-\tau/\textrm{T}_2\right)\right] - \frac{1}{4}\left[1 - \exp\left(-\tau/\textrm{T}_1\right)\right]\\
    \end{split}
    \end{align}

    \noindent where $\textrm{T}_1$ is the longitudinal relaxation time and $\textrm{T}_2$ is the dephasing time. To ensure measurement errors in the time-dependent model match the probability $p_\textrm{B} = p$ as seen in the SI1000 model, we parametrise the measurement bit-flip channel probability via

    \begin{equation}\label{eq:idle_meas_error}
        p_\textrm{B}(\tau_{\textrm{M}}) = 1 - \exp\left(-\tau_{\textrm{M}}/\tau_\textrm{D}\right)
    \end{equation}

    \noindent where $\tau_\textrm{D}$ is a characteristic depolarising timescale of the device. We choose $\tau_\textrm{D}$ such that the measurement bit-flip probability $p_B$ in the time-dependent noise model results in an error rate of $p$ given a reference measurement time $\tau_{\textrm{M}}=500~\text{ns}$.
    
    For simplicity, we assume that the measurement response of every qubit is identical, i.e. the parameters $\text{T}_1$ and $\tau_{\textrm{F}}$ are uniform across all qubits in the device. This is a limitation of our model, and in the real world manufacturing inconsistencies may lead to non-uniform measurement characteristics for different qubits. As a consequence, optimal readout fidelity may be achieved with different readout durations $\tau_{\textrm{M}}$ based on the properties of each qubit. Future investigations may illuminate how non-uniform qubit measurement responses affect the effectiveness of the soft decoding techniques outlined in this work.
    
    \subsubsection{Neutral atom noise model and  simulations}\label{methods:neutral_atom_noise_model} 

    For neutral atoms, we use a $Z$-biased circuit-level noise model, as the prevalent errors in neutral atoms are dephasing errors \cite{Cong2022FTQC,Evered2023CZfidelityNA,chamberland2022universal, higgott2023improved}. We carry out quantum memory simulations on a rotated planar surface code for neutral atoms, for ease of comparison with the superconducting qubit simulations. The physical noise parameters for neutral atoms are based on Ref.~\cite{wintersperger2023neutral, Cong2022FTQC, Evered2023CZfidelityNA}: single-qubit gate duration of $\tau_{\textrm{1q}}\!\!=\!\!500~\textrm{ns}$, two-qubit gate (CZ) duration of $\tau_{\textrm{2q}}\!\!=\!\!270~\textrm{ns}$, and $\tau_{\textrm{R}}\!\!=\!\!2000~\textrm{ns}$ for reset in $Z$-basis. The dominant sources of error in neutral atoms are due to two-qubit gates implemented using Rydberg blockade. The single-qubit and idling errors in neutral atom platforms are insignificant due to the long coherence times \cite{Wu2022erasure}, therefore, we include only a minor amount of depolarising error with probability $p/10$ as the idling error. We introduce time-dependence into our noise model by changing the measurement time, as the readout probability distribution directly depends on $\tau_{\textrm{M}}$ as seen in \cref{eq:NA_pdf}.
    
    An essential feature to enable large-scale quantum simulations with neutral atoms is simultaneous and non-destructive readout of many atomic qubits. To enable soft-information decoding, our readout noise model is based on non-destructive state selective readout via fluorescence detection \cite{MartinezDorantes2018state-dependent, radnaev2024universal, finkelstein2024universal}. For neutral atoms such as $^{87}\textrm{Rb}$, the readout scheme distinguishes between two hyperfine levels, $F=1$ and $F=2$ of the $5S_{1/2}$ ground state using fluorescence detection. In this work we simulate readout via a single photon counting module (SPCM) as proposed in \cite{shea2018fast,Deist2022midcktNA}, with the soft measurement value $\mu$ representing the photon count.

    As the atom is illuminated by the probe beam, one state appears bright (scatters photons) and the other state appears dark (no photons are scattered). For ideal photon detection, if the number of photons detected is greater or equal to a set threshold photon number $\mu_{\textrm{th}}$ the atom is said to be in bright state. If the number of detected photons is less than $\mu_{\textrm{th}}$, the atom is said to be in dark state. However, in reality the photon detectors have background noise or dark counts which can lead to misclassification of the state the atom is in, especially when the atom is in dark state. Additionally, there exists inherent uncertainty in the number of photons detected due to the Poisson distribution of the counts. The bright state errors arise due to experimental imperfections, state preparation, and inherent physical processes. \cref{eq:NA_pdf}, taken from Ref.~\cite{shea2018fast} describes the probability density functions $f^{(\bar{\mu}=0)}(\mu)$ and $f^{(\bar{\mu}=1)}(\mu)$ corresponding to the dark and the bright states respectively.

    \begin{widetext}
    \begin{subequations} \label{eq:NA_pdf}
    \begin{align}
    f^{(\bar{\mu}=1)}(\mu;t) &= e^{-(\eta R_0 + R_\textrm{bg} + R_{b \to d})t} \frac{(\eta R_0 + R_\textrm{bg})^\mu t^\mu}{\mu!} 
    + \left( \frac{R_{b \to d} e^{-R_\textrm{bg} t}}{\eta R_0 + R_{b \to d}} \right)
    \left( \frac{\eta R_0}{\eta R_0 + R_{b \to d}} \right)^\mu \notag \\
    &\quad \times \left[ \sum_{k=0}^{\mu} 
    \frac{(\eta R_0 + R_{b \to d})^k (R_\textrm{bg} t)^k}{k!(\eta R_0)^k}
    - e^{-(\eta R_0 + R_{b \to d})t} \sum_{k=0}^{n} 
    \frac{(\eta R_0 + R_{b \to d})^k (\eta R_0 + R_\textrm{bg})^k t^k}{k!(\eta R_0)^k} \right] 
    \end{align}
    \begin{align}
    f^{(\bar{\mu}=0)}(\mu;t) &= e^{-(R_\textrm{bg} + R_{d \to b})t} \frac{(R_\textrm{bg}t)^\mu}{\mu!} 
    + \left( \frac{R_{d \to b} e^{-R_\textrm{bg} t}}{\eta R_0 - R_{d \to b}} \right)
    \left( \frac{\eta R_0}{\eta R_0 - R_{d \to b}} \right)^\mu \notag \\
    &\quad \times \left[ e^{-R_{d \to b} t} \sum_{k=0}^{\mu} 
    \frac{(\eta R_0 - R_{d \to b})^k (R_\textrm{bg} t)^k}{k!(\eta R_0)^k}
    - e^{-\eta R_0 t} \sum_{k=0}^{\mu} 
    \frac{(\eta R_0 - R_{d \to b})^k (\eta R_0 + R_\textrm{bg})^k t^k}{k!(\eta R_0)^k} \right] 
    \end{align}
    \end{subequations}
    \end{widetext}

    \noindent The main parameters for this readout model include the scattering rate of an atom in the bright state $R_{0}$, the detection efficiency $\eta$, the background scattering rate $R_{\textrm{bg}}$, the scattering rate when atoms transition from a bright to a dark state $R_{b \to d}$, and the rate of scattering when atoms transition from a dark to a bright state $R_{d \to b}$. In a setting where background scattering and transition errors are low, the longer the readout pulse the smaller the overlap is between the two distributions, resulting in a lower soft measurement flip probability. The distributions for the bright and the dark states are Poissonian, with an effective mean no. of photons $\lambda_{\textrm{bright}}=(\eta R_{0}+R_{\textrm{bg}})\tau_{\textrm{M}}$ in a bright state and $\lambda_{\textrm{dark}}=R_{\textrm{bg}}\tau_{\textrm{M}}$ in a dark state. To reflect a realistic neutral atom readout chain \cite{shea2018fast}, our simulations are configured with $\eta=0.1\%$, $\tau_\textrm{M}=100~\mu\textrm{s}$, $R_0=10^7$, $R_{\textrm{bg}}=10^3$, $R_{b \to d}=960$ and $R_{d \to b}=2$. The qubits in neutral atoms are naturally identical and have long coherence times, therefore, it is safe to assume uniform measurement response functions across all qubits in a simulation.

    \subsection{Estimating the error suppression rate}\label{methods:lambda_fitting}

    The following process describes how we extract the error suppression rate $\Lambda$ from surface code simulations of variable code distances $d$. The method is applicable to experiments in both $X$- and $Z$-basis. When quoting results for a $\Lambda$-factor in \cref{results}, we take the average across $\Lambda_X$ and $\Lambda_Z$.
    
    In a quantum memory experiment taking $T$ rounds of syndrome extraction, the logical error probability is given by \cref{eq:p_l_per_round}. By repeatedly sampling and decoding syndromes from the circuit, we obtain an estimate of $P_L$. Solving \cref{eq:p_l_per_round} for the per-round error rate $\epsilon_d$ we get

    \begin{equation}\label{eq:epsilon_d_from_pl}
        \epsilon_d = \frac{1}{2}\left(1 - \left(1 - 2P_L\right)^{1/T}\right)
    \end{equation}

    \noindent where we substitute the measured value of $P_L$. To find $\Lambda$, we first compute the per-round error rate $\epsilon_d$ from \cref{eq:lambda} for varying code distances $d$, and take a least-squares fit of $x=(d + 1)/2$ against $y=\log\left(\epsilon_d\right)$. The resulting fit takes the form

    \begin{equation}\label{eq:log_e_L}
        \log\left(\epsilon_d\right) = -\log(\Lambda)\times d + \log(p_0)
    \end{equation}
    
    \noindent where $p_0$ is a constant offset, and we can extract $\Lambda$ from the coefficient in front of $d$. To estimate the uncertainty in $\Lambda$, we use the Sinter library (a sub-package of Stim \cite{gidney2021stim}) to evaluate hypotheses with a likelihood within a factor of 10 of the maximum likelihood hypothesis, given the sampled data. This method gives us error bars on the logical error probability $P_{\textrm{L}}$, avoiding the pitfall of making overconfident estimates about the accuracy of the logical error probability at low error rates. We propagate the Sinter-derived uncertainty $\Delta P_{\textrm{L}}$ to $\epsilon_d$ via \cref{eq:epsilon_d_from_pl} and use the resulting error bars in the curve fitting process of \cref{eq:log_e_L} to determine the standard error in $\Lambda$.
    
    \subsection{Soft information for LCD}\label{methods:graph_decoding}

    Our method for incorporating soft information into the local clustering decoder follows the process outlined by Pattison \textit{et al.} in Ref.~\cite{pattison2021improved}. When performing QEC with the surface code, stabiliser measurements are repeated to identify potential errors in the system \cite{fowler2012surface}. By choosing sets of stabiliser measurements whose outcome is deterministic in the absence of noise, referred to as \textit{detectors} \cite{gidney2021stim}, we can infer the effect of errors on the state of the logical qubit. For the surface code, it is possible to represent the decoding problem in a \textit{decoding graph}, where the nodes are formed of detectors, connected by edges of possible error mechanisms. The task of the classical decoder is to identify sets of edges (or clusters of edges) that connect together every detection event. To account for the different probabilities $p_i$ of $i$ distinct error mechanisms in the system, edges in the graph $e_i$ can have non-uniform weights $w$ given by
    \begin{equation}\label{eq:edge_w_given_p}
        w(e_i) = -\log\left(\frac{p_i}{1 - p_i}\right) \quad.
    \end{equation}

    By introducing weights to the graph, it is possible to find the most likely error set $\{e_i\}$ that explains the observed detection events. To do this, one needs to find a minimum-weight perfect matching in the graph, i.e. a set of errors $\{e_i\}$ that minimises the error probability $\sum_i P(e_i)$ while satisfying the observed syndrome. The more likely an error, the lower its weight, and a minimum-weight perfect matching decoder is able to effectively find a solution to this problem in polynomial time \cite{higgott2023sparse}.
    
    Instead of MWPM, in this work we use the almost-linear-time Union Find (UF) algorithm which approximates the minimum-weight perfect matching problem, as proposed in Ref.~\cite{delfosse2021almost}. The decoding algorithm is implemented in the local clustering decoder~\cite{ziad2025local}. Instead of solving the matching problem, UF solves a clustering problem, making it slightly less accurate for the purposes of decoding but considerably faster to implement in hardware. Despite the marginal accuracy penalty, LCD has similar sub-threshold scaling to MWPM and its architecture is designed for fast on-the fly edge updates, making it well suited for the adaptive graph modifications required for soft information decoding.
    
    To introduce soft information to the decoding graph, we dynamically update the edge weights $w(e_j)$ corresponding to the $M$ measurement error mechanisms $e_j$ for $j=\{1, 2, ..., M\}$ according to the posterior probabilities $P(1 \mid \mu)$ as defined in \cref{eq:post_1}. We repeat the process for every shot in the experiment, ensuring that the edge weights in the decoding graph accurately represent the probabilities of measurement errors happening in the system. The measurement edge weights are updated according to \cref{eq:edge_w_given_p} where the new error probability $p_i \rightarrow p_i^\prime$ of an edge is given by
    \begin{equation}\label{eq:p_update_rule}
        p_i^\prime = p_{S, i}\left(1 - p_i\right) + p_i\left(1 - p_{S, i}\right)
    \end{equation}
    where $p_{S, i} = \min\left[P(1 \mid \mu_i), 1 - P(1 \mid \mu_i)\right]$ is the probability of a soft measurement flip. In the case where the pre-existing edge probability $p_i$ of an error mechanism is zero, we can directly substitute $p_i^\prime = p_{S, i}$ into \cref{eq:edge_w_given_p}. Once we have a soft-information-adjusted decoding graph, we can decode it as usual. To ensure fast decoding in our simulations, we truncate the measurement probabilities into $8$ bits per measurement, restricting the number of possible graph updates needed. To compare decoding performance between a soft and a hard decoder in the presence of classification errors, we introduce an additional error component into the decoding graph of the hard decoder based on the shot-averaged soft error probability $p_\textrm{S}$. This technique, denoted \textit{data-informed hard matching} in Ref.~\cite{hanisch2024soft} ensures that the comparison between the soft and the hard decoder is fair.
    
    \subsection{Soft Belief Propagation Decoder}\label{methods: softBP}        
    Belief Propagation (BP), also known as sum-product algorithm, is a message-passing algorithm used in solving classical inference problems \cite{Kschischang2001productsum,noorshams2013productsum}. For classical BP decoding, the messages are passed along the edges of a Tanner graph (or a factor graph). In the classical decoding problem, the Tanner graph---not to be confused with the decoding graph of the previous subsection---is a bipartite graph composed of check nodes and variable nodes (physical bits). There exists an edge between a check node and a variable node if that error can change the parity of the check nodes. The error nodes contain the information about the error mechanisms present, as provided by an underlying noise model. Given the decoding problem $s=H\cdot e$ where $H$ is the parity check matrix, $s$ is the syndrome and $e$ is the error string, BP obtains the bitwise most likely error as shown in \cref{eq:error_bit} in polynomial time.
    
    \begin{align}
    \label{eq:error_bit}
        e_{i}^{\textrm{min-weight}}=\textrm{arg max}_{e_{i}}\sum_{e_{1},e_{2}...e_{n}}P(e_{1},e_{2}...e_{n}|s).
    \end{align} 
    
    \begin{figure}[h]
    \centering
    \includegraphics[width=0.85\columnwidth]{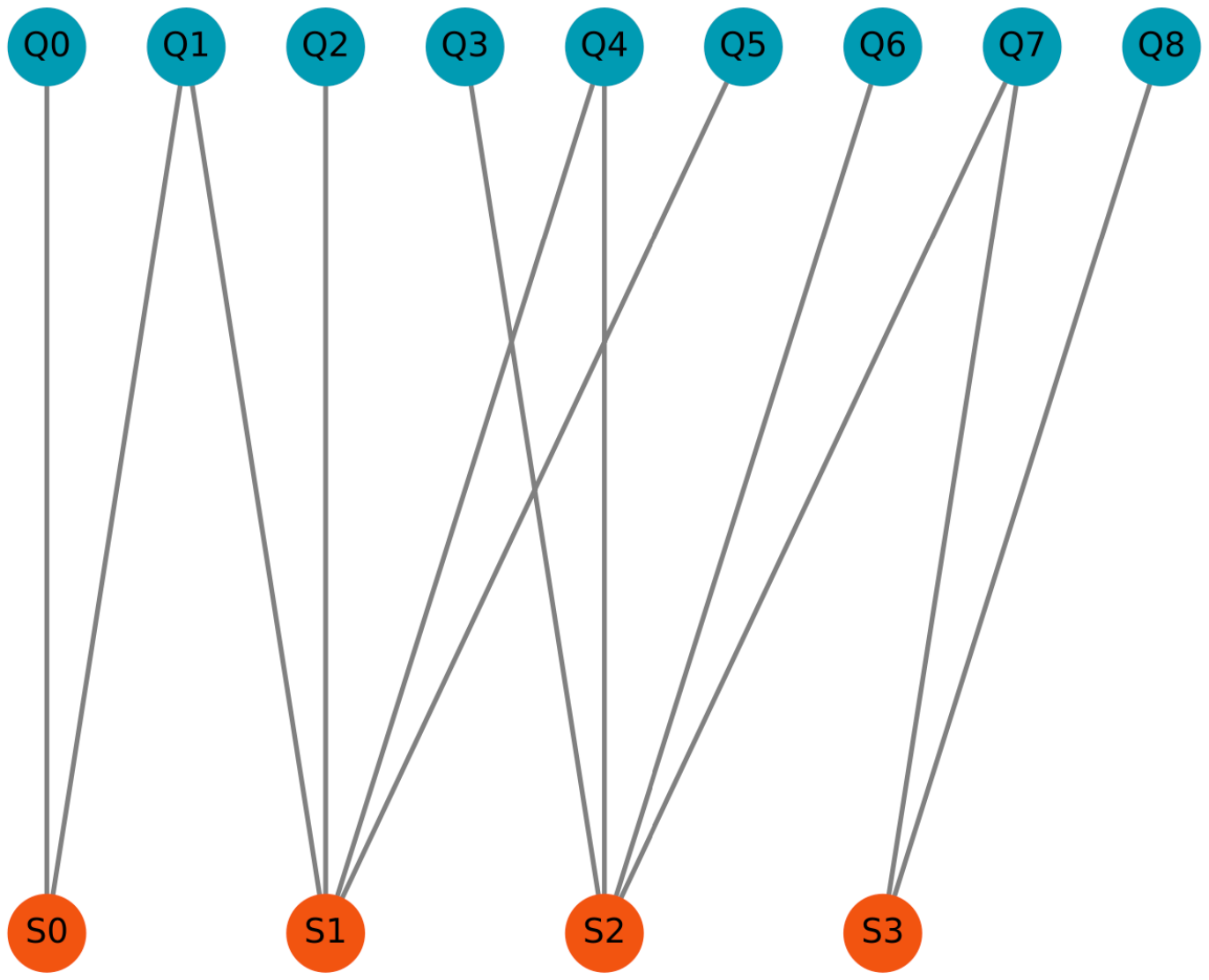}
    \caption{\textbf{Tanner graph for $X$-checks in one round of a quantum memory experiment on a distance-3 surface code}. The variable nodes that contain the error mechanisms in the syndrome extraction circuit are represented by the $Q_{i}$ nodes and the stabilizer nodes (also called check nodes) are represented by the $S_{i}$ nodes. In this case we show the X-type stabilizer checks. The edges are drawn between qubits involved in each X-type stabilizer operation.}
    \label{fig:bp_tanner_graph}
    \end{figure}

    For our purposes, we focus on the Calderbank-Shor-Steane (CSS) subclass of qLDPC codes under a circuit-level noise model. We show an example of the Tanner graph for a distance-$3$ surface code in \cref{fig:bp_tanner_graph}. The Tanner graph for a circuit-level noise model can be represented as $\mathcal{T}=(Q,S,E)$ containing a set of variables nodes ($Q_{i}$) which store the information about the error mechanisms based on the noise model used, while the check nodes ($S_{i}$) correspond to the detectors and $E$ is the edge set. If the error mechanism flips the detectors there exists an edge between $Q_{i}$ and $S_{i}$.  In the case of multiple error mechanisms flipping the same detectors, we merge these error mechanisms into one variable node---i.e. merge the equivalent variable nodes.  The BP decoder is initialized with log-likelihood ratios, also known as \textit{priors}, the probabilities of which we obtain from the circuit-level noise model. The use of log-likelihood ratios is to ensure numerical stability of BP.

    Next, we describe the soft information enhanced BP-based decoding algorithm. Given the Tanner graph $\mathcal{T}=(Q,S,E)$ the prior probabilities for each variable node $Q_{i}$ contains the probability that it would flip the corresponding detector $S_{i}$. Previous works \cite{berent2024analog, Grospellier2021combininghardsoft} add soft information to the Tanner graph by introducing additional variable nodes to the graph. These virtual nodes ensure that the measurement errors are correctly tracked. Also, in the absence of resets, measurement errors can propagate to future rounds giving rise to time-like edges, so adding virtual nodes to the Tanner graph helps capture these errors as they propagate over time. In this work we show how to add soft information to Tanner graph without adding virtual nodes to it.

    \begin{algorithm}
        \caption{Soft-Belief Propagation}
        \label{alg:soft-BP}

    \textbf{Input:} Circuit-level Tanner graph $\mathcal{T}_{\textrm{circ}}$, posterior measurement probabilities $P(1\mid\mu_i)$ for each soft measurement $\mu_i$, prior probabilities ${p_{\textrm{priors}}}$ .\\
     
    \textbf{Output:} Variable nodes in $\mathcal{T}_{\textrm{circ}}$ corresponding to the (hyper)edges updated with $p_{\textrm{S}}$.
        \begin{algorithmic}[1]
          \State Compute a mapping of  detectors $S_{i}$ and measurement errors indices.
          \State Retrieve prior probabilities from the error model, prior($Q_{i}$).
          \State Compute soft flip probabilities $p_{S,i}$ from $P(1\mid\mu_i)$ using \cref{eq:p-soft}. 
          \State Compute total error probability of each prior $Q_i$ using \cref{eq:p_update_rule}.
          \State Initialize the BP decoder with new $p_{\textrm{priors}}$.        
        \end{algorithmic}
        \textbf{Return:} $Q_{i}$.
    \end{algorithm}
    
    In order to include classification error probabilities derived from the soft information model, we need to modify the prior probabilities. As mentioned earlier, the variable nodes contain the physical error mechanisms and measurement errors. We first separate the measurement errors from the rest of the errors, by mapping the index of the measurement errors $i=\{1,2,...,n_{\textrm{meas}}\}$ where $n_{\textrm{meas}}$ is the total number of measurement errors in the circuit to the set of detectors $\{S_{i}\}$ they trigger. In the case of a syndrome extraction circuit that contains resets after measurements, measurement classification errors trigger the same detectors as pre-measurement bit-flip errors---meaning that we do not need to introduce additional variable nodes into the Tanner graph. Now, we can dynamically update the prior probabilities of the the variable nodes $Q_{i}$ with the soft measurement error probabilities using \cref{eq:p_update_rule}. The BP decoder is then initialized with new log-likelihood ratios obtained form these updated prior probabilities. In cases where the same variable node $Q_{i}$ is affected by several different measurements, we follow the process recursively, until all the soft measurements have been incorporated into the priors. This process is described in Algorithm~\ref{alg:soft-BP}.
    
    \noindent Given the updated priors, we then decode the syndrome using BP+OSD for the case of bivariate bicycle (BB) codes and BP-matching on surface codes. The process of updating the priors is repeated for every shot.

\section{Data and code availability}

    The Stim circuits used to simulate the experiments in this study, as well as code used to generate soft measurement samples for superconducting and neutral atom platforms are available at \url{https://github.com/riverlane/soft_information_models}. Additional data used to support the findings of this study can be made available by the corresponding authors upon reasonable request.

\section{Acknowledgments}

    The authors would like to thank Joan Camps and Neil Gillespie for insightful discussions during the project, as well as their feedback on the manuscript. The authors would also like to thank Earl Campbell for his input at various stages of the work, and Ophelia Crawford for her contributions to the early stages of the research. 

\section{Author contributions}

    JM led the research and writing of the manuscript, modelled soft information in the context of superconducting qubits, incorporated soft incorporation decoding into the local clustering decoder and developed software to simulate and analyse decoding performance under varying noise regimes. ESM led the implementation of soft-information-aware belief propagation algorithm, designed and performed simulations of QEC experiments with the soft belief propagation decoder and modelled QEC experiments on neutral atom platforms. The manuscript was jointly written by both authors.

\section{Competing Interests}

    The authors declare no competing interests.

\section{Copyright}

    This is the Accepted Manuscript version of an article accepted for publication in Quantum Science and Technology. IOP Publishing Ltd is not responsible for any errors or omissions in this version of the manuscript or any version derived from it. This Accepted Manuscript is published under a \href{https://creativecommons.org/licenses/by/4.0/}{CC BY} licence. The Version of Record is available online at \url{https://doi.org/10.1088/2058-9565/ae4e55}.

\bibliography{arxiv_v3}

\begin{thebibliography}{10}
\expandafter\ifx\csname url\endcsname\relax
  \def\url#1{\texttt{#1}}\fi
\expandafter\ifx\csname urlprefix\endcsname\relax\def\urlprefix{URL }\fi
\providecommand{\bibinfo}[2]{#2}
\providecommand{\eprint}[2][]{\url{#2}}

\bibitem{google2024quantum}
\bibinfo{author}{{Google Quantum AI}}.
\newblock \bibinfo{title}{Quantum error correction below the surface code threshold}.
\newblock \emph{\bibinfo{journal}{Nature}} \textbf{\bibinfo{volume}{638}}, \bibinfo{pages}{920--926} (\bibinfo{year}{2025}).

\bibitem{bluvstein2024logical}
\bibinfo{author}{Bluvstein, D.} \emph{et~al.}
\newblock \bibinfo{title}{Logical quantum processor based on reconfigurable atom arrays}.
\newblock \emph{\bibinfo{journal}{Nature}} \textbf{\bibinfo{volume}{626}}, \bibinfo{pages}{58--65} (\bibinfo{year}{2024}).

\bibitem{paetznick2024demonstration}
\bibinfo{author}{Paetznick, A.} \emph{et~al.}
\newblock \bibinfo{title}{Demonstration of logical qubits and repeated error correction with better-than-physical error rates}.
\newblock \emph{\bibinfo{journal}{arXiv}}  (\bibinfo{year}{2024}).
\newblock \urlprefix\url{https://arxiv.org/abs/2404.02280}.

\bibitem{kitaev2003fault-tolerant}
\bibinfo{author}{Kitaev, A.}
\newblock \bibinfo{title}{Fault-tolerant quantum computation by anyons}.
\newblock \emph{\bibinfo{journal}{Annals of Physics}} \textbf{\bibinfo{volume}{303}}, \bibinfo{pages}{2--30} (\bibinfo{year}{2003}).

\bibitem{aharonov2008fault-tolerant}
\bibinfo{author}{Aharonov, D.} \& \bibinfo{author}{Ben-Or, M.}
\newblock \bibinfo{title}{Fault-tolerant quantum computation with constant error rate}.
\newblock \emph{\bibinfo{journal}{SIAM Journal on Computing}} \textbf{\bibinfo{volume}{38}}, \bibinfo{pages}{1207--1282} (\bibinfo{year}{2008}).

\bibitem{beverland2022assessing}
\bibinfo{author}{Beverland, M.~E.} \emph{et~al.}
\newblock \bibinfo{title}{Assessing requirements to scale to practical quantum advantage}.
\newblock \emph{\bibinfo{journal}{arXiv}}  (\bibinfo{year}{2022}).
\newblock \urlprefix\url{https://arxiv.org/abs/2211.07629}.

\bibitem{mohseni2025build}
\bibinfo{author}{Mohseni, M.} \emph{et~al.}
\newblock \bibinfo{title}{How to build a quantum supercomputer: Scaling from hundreds to millions of qubits}.
\newblock \emph{\bibinfo{journal}{arXiv}}  (\bibinfo{year}{2025}).
\newblock \urlprefix\url{https://arxiv.org/abs/2411.10406}.

\bibitem{bausch2024learning}
\bibinfo{author}{Bausch, J.} \emph{et~al.}
\newblock \bibinfo{title}{Learning high-accuracy error decoding for quantum processors}.
\newblock \emph{\bibinfo{journal}{Nature}} \textbf{\bibinfo{volume}{635}}, \bibinfo{pages}{834--840} (\bibinfo{year}{2024}).

\bibitem{dennis2002topological}
\bibinfo{author}{Dennis, E.}, \bibinfo{author}{Kitaev, A.}, \bibinfo{author}{Landahl, A.} \& \bibinfo{author}{Preskill, J.}
\newblock \bibinfo{title}{Topological quantum memory}.
\newblock \emph{\bibinfo{journal}{Journal of Mathematical Physics}} \textbf{\bibinfo{volume}{43}}, \bibinfo{pages}{4452--4505} (\bibinfo{year}{2002}).

\bibitem{google2023suppressing}
\bibinfo{author}{{Google Quantum AI}}.
\newblock \bibinfo{title}{Suppressing quantum errors by scaling a surface code logical qubit}.
\newblock \emph{\bibinfo{journal}{Nature}} \textbf{\bibinfo{volume}{614}}, \bibinfo{pages}{676--681} (\bibinfo{year}{2023}).

\bibitem{anjou2021generalised}
\bibinfo{author}{D'Anjou, B.}
\newblock \bibinfo{title}{Generalized figure of merit for qubit readout}.
\newblock \emph{\bibinfo{journal}{Phys. Rev. A}} \textbf{\bibinfo{volume}{103}}, \bibinfo{pages}{042404} (\bibinfo{year}{2021}).

\bibitem{pattison2021improved}
\bibinfo{author}{Pattison, C.~A.}, \bibinfo{author}{Beverland, M.~E.}, \bibinfo{author}{da~Silva, M.~P.} \& \bibinfo{author}{Delfosse, N.}
\newblock \bibinfo{title}{Improved quantum error correction using soft information}.
\newblock \emph{\bibinfo{journal}{arXiv}}  (\bibinfo{year}{2021}).
\newblock \urlprefix\url{https://arxiv.org/abs/2107.13589}.

\bibitem{ali2024reducing}
\bibinfo{author}{Ali, H.} \emph{et~al.}
\newblock \bibinfo{title}{Reducing the error rate of a superconducting logical qubit using analog readout information}.
\newblock \emph{\bibinfo{journal}{Phys. Rev. Appl.}} \textbf{\bibinfo{volume}{22}}, \bibinfo{pages}{044031} (\bibinfo{year}{2024}).

\bibitem{caune2024demonstrating}
\bibinfo{author}{Caune, L.} \emph{et~al.}
\newblock \bibinfo{title}{Demonstrating real-time and low-latency quantum error correction with superconducting qubits}.
\newblock \emph{\bibinfo{journal}{arXiv}}  (\bibinfo{year}{2024}).
\newblock \urlprefix\url{https://arxiv.org/abs/2410.05202}.

\bibitem{Xue2020repetitive}
\bibinfo{author}{Xue, X.} \emph{et~al.}
\newblock \bibinfo{title}{Repetitive quantum nondemolition measurement and soft decoding of a silicon spin qubit}.
\newblock \emph{\bibinfo{journal}{Physical Review X}} \textbf{\bibinfo{volume}{10}} (\bibinfo{year}{2020}).

\bibitem{hanisch2024soft}
\bibinfo{author}{Hanisch, M.~D.}, \bibinfo{author}{Hetényi, B.} \& \bibinfo{author}{Wootton, J.~R.}
\newblock \bibinfo{title}{Soft information decoding with superconducting qubits}.
\newblock \emph{\bibinfo{journal}{arXiv}}  (\bibinfo{year}{2024}).
\newblock \urlprefix\url{https://arxiv.org/abs/2411.16228}.

\bibitem{ziad2025local}
\bibinfo{author}{Ziad, A.~B.} \emph{et~al.}
\newblock \bibinfo{title}{Local clustering decoder as a fast and adaptive hardware decoder for the surface code}.
\newblock \emph{\bibinfo{journal}{Nature Communications}} \textbf{\bibinfo{volume}{16}} (\bibinfo{year}{2025}).
\newblock \urlprefix\url{http://dx.doi.org/10.1038/s41467-025-66773-x}.

\bibitem{shen2020enhanced}
\bibinfo{author}{Shen, Y.} \emph{et~al.}
\newblock \bibinfo{title}{Enhanced belief propagation decoder for 5g polar codes with bit-flipping}.
\newblock \emph{\bibinfo{journal}{IEEE Transactions on Circuits and Systems II: Express Briefs}} \textbf{\bibinfo{volume}{67}}, \bibinfo{pages}{901--905} (\bibinfo{year}{2020}).

\bibitem{borwankar2020low}
\bibinfo{author}{Borwankar, S.} \& \bibinfo{author}{Shah, D.}
\newblock \bibinfo{title}{Low density parity check code (ldpc codes) overview}.
\newblock \emph{\bibinfo{journal}{arXiv}}  (\bibinfo{year}{2020}).
\newblock \urlprefix\url{https://arxiv.org/abs/2009.08645}.

\bibitem{muller2025improved}
\bibinfo{author}{Müller, T.} \emph{et~al.}
\newblock \bibinfo{title}{Improved belief propagation is sufficient for real-time decoding of quantum memory}.
\newblock \emph{\bibinfo{journal}{arXiv}}  (\bibinfo{year}{2025}).
\newblock \urlprefix\url{https://arxiv.org/abs/2506.01779}.
\newblock \eprint{2506.01779}.

\bibitem{radnaev2024universal}
\bibinfo{author}{Radnaev, A.~G.} \emph{et~al.}
\newblock \bibinfo{title}{A universal neutral-atom quantum computer with individual optical addressing and non-destructive readout}.
\newblock \emph{\bibinfo{journal}{arXiv}}  (\bibinfo{year}{2024}).
\newblock \urlprefix\url{https://arxiv.org/abs/2408.08288}.

\bibitem{xu2024constant-overhead}
\bibinfo{author}{Xu, Q.} \emph{et~al.}
\newblock \bibinfo{title}{Constant-overhead fault-tolerant quantum computation with reconfigurable atom arrays}.
\newblock \emph{\bibinfo{journal}{Nature Physics}} \textbf{\bibinfo{volume}{20}}, \bibinfo{pages}{1084--1090} (\bibinfo{year}{2024}).

\bibitem{bravyi2024high-threshold}
\bibinfo{author}{Bravyi, S.}, \bibinfo{author}{Cross, J., A.W.~Gambetta}, \bibinfo{author}{Maslov, D.}, \bibinfo{author}{Rall, P.} \& \bibinfo{author}{Yoder, T.~J.}
\newblock \bibinfo{title}{High-threshold and low-overhead fault-tolerant quantum memory}.
\newblock \emph{\bibinfo{journal}{Nature}} \textbf{\bibinfo{volume}{627}}, \bibinfo{pages}{778--782} (\bibinfo{year}{2024}).

\bibitem{higgott2023sparse}
\bibinfo{author}{Higgott, O.} \& \bibinfo{author}{Gidney, C.}
\newblock \bibinfo{title}{Sparse blossom: correcting a million errors per core second with minimum-weight matching}.
\newblock \emph{\bibinfo{journal}{arXiv}}  (\bibinfo{year}{2023}).
\newblock \urlprefix\url{https://arxiv.org/abs/2303.15933}.

\bibitem{varbanov2025neural}
\bibinfo{author}{Varbanov, B.~M.}, \bibinfo{author}{Serra-Peralta, M.}, \bibinfo{author}{Byfield, D.} \& \bibinfo{author}{Terhal, B.~M.}
\newblock \bibinfo{title}{Neural network decoder for near-term surface-code experiments}.
\newblock \emph{\bibinfo{journal}{Physical Review Research}} \textbf{\bibinfo{volume}{7}} (\bibinfo{year}{2025}).
\newblock \urlprefix\url{http://dx.doi.org/10.1103/PhysRevResearch.7.013029}.

\bibitem{beni2025tesseract}
\bibinfo{author}{Beni, L.~A.}, \bibinfo{author}{Higgott, O.} \& \bibinfo{author}{Shutty, N.}
\newblock \bibinfo{title}{Tesseract: A search-based decoder for quantum error correction} (\bibinfo{year}{2025}).
\newblock \urlprefix\url{https://arxiv.org/abs/2503.10988}.
\newblock \eprint{2503.10988}.

\bibitem{kelly2015state}
\bibinfo{author}{Kelly, J.} \emph{et~al.}
\newblock \bibinfo{title}{State preservation by repetitive error detection in a superconducting quantum circuit}.
\newblock \emph{\bibinfo{journal}{Nature}} \textbf{\bibinfo{volume}{519}}, \bibinfo{pages}{66--69} (\bibinfo{year}{2015}).

\bibitem{shea2018fast}
\bibinfo{author}{Shea, M.~E.}
\newblock \emph{\bibinfo{title}{Fast, Nondestructive Quantum-state Readout of a Single, Trapped, Neutral Atom}}.
\newblock \bibinfo{type}{Ph.d. thesis}, \bibinfo{school}{Duke University} (\bibinfo{year}{2018}).

\bibitem{krantz2019quantum}
\bibinfo{author}{Krantz, P.} \emph{et~al.}
\newblock \bibinfo{title}{A quantum engineer's guide to superconducting qubits}.
\newblock \emph{\bibinfo{journal}{Applied Physics Reviews}} \textbf{\bibinfo{volume}{6}}, \bibinfo{pages}{021318} (\bibinfo{year}{2019}).

\bibitem{crain2016integrated}
\bibinfo{author}{Crain, S.~G.}
\newblock \emph{\bibinfo{title}{Integrated System Technologies for Modular Trapped Ion Quantum Information Processing}}.
\newblock \bibinfo{type}{Ph.d. thesis}, \bibinfo{school}{Duke University} (\bibinfo{year}{2016}).

\bibitem{noek2013high}
\bibinfo{author}{Noek, R.} \emph{et~al.}
\newblock \bibinfo{title}{High speed, high fidelity detection of an atomic hyperfine qubit}.
\newblock \emph{\bibinfo{journal}{Opt. Lett.}} \textbf{\bibinfo{volume}{38}}, \bibinfo{pages}{4735--4738} (\bibinfo{year}{2013}).

\bibitem{MartinezDorantes2018state-dependent}
\bibinfo{author}{Martinez-Dorantes, M.} \emph{et~al.}
\newblock \bibinfo{title}{State-dependent fluorescence of neutral atoms in optical potentials}.
\newblock \emph{\bibinfo{journal}{Phys. Rev. A}} \textbf{\bibinfo{volume}{97}}, \bibinfo{pages}{023410} (\bibinfo{year}{2018}).

\bibitem{heinsoo2018rapid}
\bibinfo{author}{Heinsoo, J.} \emph{et~al.}
\newblock \bibinfo{title}{Rapid high-fidelity multiplexed readout of superconducting qubits}.
\newblock \emph{\bibinfo{journal}{Phys. Rev. Appl.}} \textbf{\bibinfo{volume}{10}}, \bibinfo{pages}{034040} (\bibinfo{year}{2018}).

\bibitem{bengtsson2024model-based}
\bibinfo{author}{Bengtsson, A.} \emph{et~al.}
\newblock \bibinfo{title}{Model-based optimization of superconducting qubit readout}.
\newblock \emph{\bibinfo{journal}{Phys. Rev. Lett.}} \textbf{\bibinfo{volume}{132}}, \bibinfo{pages}{100603} (\bibinfo{year}{2024}).

\bibitem{Cong2022FTQC}
\bibinfo{author}{Cong, I.} \emph{et~al.}
\newblock \bibinfo{title}{Hardware-efficient, fault-tolerant quantum computation with rydberg atoms}.
\newblock \emph{\bibinfo{journal}{Phys. Rev. X}} \textbf{\bibinfo{volume}{12}}, \bibinfo{pages}{021049} (\bibinfo{year}{2022}).

\bibitem{geher2025toreset}
\bibinfo{author}{Geh{\'e}r, G.~P.}, \bibinfo{author}{Jastrzebski, M.}, \bibinfo{author}{Campbell, E.~T.} \& \bibinfo{author}{Crawford, O.}
\newblock \bibinfo{title}{To reset, or not to reset---that is the question}.
\newblock \emph{\bibinfo{journal}{npj Quantum Information}} \textbf{\bibinfo{volume}{11}}, \bibinfo{pages}{39} (\bibinfo{year}{2025}).
\newblock \urlprefix\url{https://doi.org/10.1038/s41534-025-00998-y}.

\bibitem{google2021exponential}
\bibinfo{author}{Chen, Z.} \emph{et~al.}
\newblock \bibinfo{title}{Exponential suppression of bit or phase errors with cyclic error correction}.
\newblock \emph{\bibinfo{journal}{Nature}} \textbf{\bibinfo{volume}{595}}, \bibinfo{pages}{383–387} (\bibinfo{year}{2021}).
\newblock \urlprefix\url{http://dx.doi.org/10.1038/s41586-021-03588-y}.

\bibitem{Note1}
\bibinfo{note}{This assumes lattice surgery as the means of logical computation. For qubit platforms such as neutral atoms that are compatible with transversal logic, only $\protect \mathcal {O}(1)$ rounds are needed for logical operations and one can achieve the same logical error rate with a lower-distance code}.

\bibitem{preskill2025megaquop}
\bibinfo{author}{Preskill, J.}
\newblock \bibinfo{title}{Beyond nisq: The megaquop machine}.
\newblock \emph{\bibinfo{journal}{ACM Transactions on Quantum Computing}} \textbf{\bibinfo{volume}{6}}, \bibinfo{pages}{1–7} (\bibinfo{year}{2025}).
\newblock \urlprefix\url{http://dx.doi.org/10.1145/3723153}.

\bibitem{gidney2021stim}
\bibinfo{author}{Gidney, C.}
\newblock \bibinfo{title}{Stim: a fast stabilizer circuit simulator}.
\newblock \emph{\bibinfo{journal}{{Quantum}}} \textbf{\bibinfo{volume}{5}}, \bibinfo{pages}{497} (\bibinfo{year}{2021}).

\bibitem{gidney2021fault-tolerant}
\bibinfo{author}{Gidney, C.}, \bibinfo{author}{Newman, M.}, \bibinfo{author}{Fowler, A.} \& \bibinfo{author}{Broughton, M.}
\newblock \bibinfo{title}{A {F}ault-{T}olerant {H}oneycomb {M}emory}.
\newblock \emph{\bibinfo{journal}{{Quantum}}} \textbf{\bibinfo{volume}{5}}, \bibinfo{pages}{605} (\bibinfo{year}{2021}).

\bibitem{piskor2025realquantum}
\bibinfo{author}{Piskor, T.} \emph{et~al.}
\newblock \bibinfo{title}{Simulation and benchmarking of real quantum hardware} (\bibinfo{year}{2025}).
\newblock \urlprefix\url{https://arxiv.org/abs/2508.04483}.
\newblock \eprint{2508.04483}.

\bibitem{Evered2023CZfidelityNA}
\bibinfo{author}{Evered, S.~J.} \emph{et~al.}
\newblock \bibinfo{title}{High-fidelity parallel entangling gates on a neutral-atom quantum computer}.
\newblock \emph{\bibinfo{journal}{Nature}} \textbf{\bibinfo{volume}{622}}, \bibinfo{pages}{268--272} (\bibinfo{year}{2023}).

\bibitem{graham2023NA}
\bibinfo{author}{Graham, T.~M.} \emph{et~al.}
\newblock \bibinfo{title}{Midcircuit measurements on a single-species neutral alkali atom quantum processor}.
\newblock \emph{\bibinfo{journal}{Phys. Rev. X}} \textbf{\bibinfo{volume}{13}}, \bibinfo{pages}{041051} (\bibinfo{year}{2023}).

\bibitem{nikolov2023RBNA}
\bibinfo{author}{Nikolov, B.}, \bibinfo{author}{Diamond-Hitchcock, E.}, \bibinfo{author}{Bass, J.}, \bibinfo{author}{Spong, N. L.~R.} \& \bibinfo{author}{Pritchard, J.~D.}
\newblock \bibinfo{title}{Randomized benchmarking using nondestructive readout in a two-dimensional atom array}.
\newblock \emph{\bibinfo{journal}{Phys. Rev. Lett.}} \textbf{\bibinfo{volume}{131}}, \bibinfo{pages}{030602} (\bibinfo{year}{2023}).

\bibitem{ghosh2012surface}
\bibinfo{author}{Ghosh, J.}, \bibinfo{author}{Fowler, A.~G.} \& \bibinfo{author}{Geller, M.~R.}
\newblock \bibinfo{title}{Surface code with decoherence: An analysis of three superconducting architectures}.
\newblock \emph{\bibinfo{journal}{Phys. Rev. A}} \textbf{\bibinfo{volume}{86}}, \bibinfo{pages}{062318} (\bibinfo{year}{2012}).

\bibitem{raveendran2022soft}
\bibinfo{author}{Raveendran, N.}, \bibinfo{author}{Rengaswamy, N.}, \bibinfo{author}{Pradhan, A.~K.} \& \bibinfo{author}{Vasić, B.}
\newblock \bibinfo{title}{Soft syndrome decoding of quantum ldpc codes for joint correction of data and syndrome errors}.
\newblock In \emph{\bibinfo{booktitle}{2022 IEEE International Conference on Quantum Computing and Engineering (QCE)}}, \bibinfo{pages}{275--281} (\bibinfo{year}{2022}).

\bibitem{berent2024analog}
\bibinfo{author}{Berent, L.}, \bibinfo{author}{Hillmann, T.}, \bibinfo{author}{Eisert, J.}, \bibinfo{author}{Wille, R.} \& \bibinfo{author}{Roffe, J.}
\newblock \bibinfo{title}{Analog information decoding of bosonic quantum low-density parity-check codes}.
\newblock \emph{\bibinfo{journal}{PRX Quantum}} \textbf{\bibinfo{volume}{5}}, \bibinfo{pages}{020349} (\bibinfo{year}{2024}).

\bibitem{higgott2023improved}
\bibinfo{author}{Higgott, O.}, \bibinfo{author}{Bohdanowicz, T.~C.}, \bibinfo{author}{Kubica, A.}, \bibinfo{author}{Flammia, S.~T.} \& \bibinfo{author}{Campbell, E.~T.}
\newblock \bibinfo{title}{Improved decoding of circuit noise and fragile boundaries of tailored surface codes}.
\newblock \emph{\bibinfo{journal}{Phys. Rev. X}} \textbf{\bibinfo{volume}{13}}, \bibinfo{pages}{031007} (\bibinfo{year}{2023}).

\bibitem{Pecorari2025}
\bibinfo{author}{Pecorari, L.}, \bibinfo{author}{Jandura, S.}, \bibinfo{author}{Brennen, G.~K.} \& \bibinfo{author}{Pupillo, G.}
\newblock \bibinfo{title}{High-rate quantum ldpc codes for long-range-connected neutral atom registers}.
\newblock \emph{\bibinfo{journal}{Nature Communications}} \textbf{\bibinfo{volume}{16}}, \bibinfo{pages}{1111} (\bibinfo{year}{2025}).

\bibitem{roffe2020decoding}
\bibinfo{author}{Roffe, J.}, \bibinfo{author}{White, D.~R.}, \bibinfo{author}{Burton, S.} \& \bibinfo{author}{Campbell, E.}
\newblock \bibinfo{title}{Decoding across the quantum low-density parity-check code landscape}.
\newblock \emph{\bibinfo{journal}{Physical Review Research}} \textbf{\bibinfo{volume}{2}} (\bibinfo{year}{2020}).

\bibitem{panteleev2021degeneratequantum}
\bibinfo{author}{Panteleev, P.} \& \bibinfo{author}{Kalachev, G.}
\newblock \bibinfo{title}{Degenerate {Q}uantum {LDPC} {C}odes {W}ith {G}ood {F}inite {L}ength {P}erformance}.
\newblock \emph{\bibinfo{journal}{{Quantum}}} \textbf{\bibinfo{volume}{5}}, \bibinfo{pages}{585} (\bibinfo{year}{2021}).

\bibitem{wolanski2024ambiguity}
\bibinfo{author}{Wolanski, S.} \& \bibinfo{author}{Barber, B.}
\newblock \bibinfo{title}{Ambiguity clustering: an accurate and efficient decoder for qldpc codes}.
\newblock \emph{\bibinfo{journal}{arXiv}}  (\bibinfo{year}{2024}).
\newblock \urlprefix\url{https://arxiv.org/abs/2406.14527}.

\bibitem{hillmann2024localized}
\bibinfo{author}{Hillmann, T.} \emph{et~al.}
\newblock \bibinfo{title}{Localized statistics decoding: A parallel decoding algorithm for quantum low-density parity-check codes}.
\newblock \emph{\bibinfo{journal}{arXiv}}  (\bibinfo{year}{2024}).
\newblock \urlprefix\url{https://arxiv.org/abs/2406.18655}.

\bibitem{gong2024lowlatencyiterativedecodingqldpc}
\bibinfo{author}{Gong, A.}, \bibinfo{author}{Cammerer, S.} \& \bibinfo{author}{Renes, J.~M.}
\newblock \bibinfo{title}{Toward low-latency iterative decoding of qldpc codes under circuit-level noise}.
\newblock \emph{\bibinfo{journal}{arXiv}}  (\bibinfo{year}{2024}).
\newblock \urlprefix\url{https://arxiv.org/abs/2403.18901}.

\bibitem{roffe2022LDPC}
\bibinfo{author}{Roffe, J.}
\newblock \bibinfo{title}{{LDPC: Python tools for low density parity check codes}} (\bibinfo{year}{2022}).
\newblock \urlprefix\url{https://pypi.org/project/ldpc/}.

\bibitem{liyanage2023scalable}
\bibinfo{author}{Liyanage, N.}, \bibinfo{author}{Wu, Y.}, \bibinfo{author}{Deters, A.} \& \bibinfo{author}{Zhong, L.}
\newblock \bibinfo{title}{Scalable quantum error correction for surface codes using fpga}.
\newblock \emph{\bibinfo{journal}{arXiv}}  (\bibinfo{year}{2023}).
\newblock \urlprefix\url{https://arxiv.org/abs/2301.08419}.

\bibitem{vora2024ml}
\bibinfo{author}{Vora, N.~R.} \emph{et~al.}
\newblock \bibinfo{title}{Ml-powered fpga-based real-time quantum state discrimination enabling mid-circuit measurements} (\bibinfo{year}{2024}).
\newblock \urlprefix\url{https://arxiv.org/abs/2406.18807}.
\newblock \eprint{2406.18807}.

\bibitem{bonilla2021xzzx}
\bibinfo{author}{Bonilla~Ataides, J.~P.}, \bibinfo{author}{Tuckett, D.~K.}, \bibinfo{author}{Bartlett, S.~D.}, \bibinfo{author}{Flammia, S.~T.} \& \bibinfo{author}{Brown, B.~J.}
\newblock \bibinfo{title}{The xzzx surface code}.
\newblock \emph{\bibinfo{journal}{Nature Communications}} \textbf{\bibinfo{volume}{12}}, \bibinfo{pages}{2172} (\bibinfo{year}{2021}).
\newblock \urlprefix\url{https://doi.org/10.1038/s41467-021-22274-1}.

\bibitem{chamberland2022universal}
\bibinfo{author}{Chamberland, C.} \& \bibinfo{author}{Campbell, E.~T.}
\newblock \bibinfo{title}{Universal quantum computing with twist-free and temporally encoded lattice surgery}.
\newblock \emph{\bibinfo{journal}{PRX Quantum}} \textbf{\bibinfo{volume}{3}}, \bibinfo{pages}{010331} (\bibinfo{year}{2022}).

\bibitem{wintersperger2023neutral}
\bibinfo{author}{Wintersperger, K.}, \bibinfo{author}{Dommert, F.}, \bibinfo{author}{Ehmer, T.} \& \bibinfo{author}{et~al.}
\newblock \bibinfo{title}{Neutral atom quantum computing hardware: performance and end-user perspective}.
\newblock \emph{\bibinfo{journal}{EPJ Quantum Technol.}} \textbf{\bibinfo{volume}{10}}, \bibinfo{pages}{21} (\bibinfo{year}{2023}).

\bibitem{Wu2022erasure}
\bibinfo{author}{Wu, Y.}, \bibinfo{author}{Kolkowitz, S.}, \bibinfo{author}{Puri, S.} \& \bibinfo{author}{Thompson, J.~D.}
\newblock \bibinfo{title}{Erasure conversion for fault-tolerant quantum computing in alkaline earth rydberg atom arrays}.
\newblock \emph{\bibinfo{journal}{Nature Communications}} \textbf{\bibinfo{volume}{13}}, \bibinfo{pages}{4657} (\bibinfo{year}{2022}).

\bibitem{finkelstein2024universal}
\bibinfo{author}{Finkelstein, R.} \emph{et~al.}
\newblock \bibinfo{title}{Universal quantum operations and ancilla-based read-out for tweezer clocks}.
\newblock \emph{\bibinfo{journal}{Nature}} \textbf{\bibinfo{volume}{634}}, \bibinfo{pages}{321--327} (\bibinfo{year}{2024}).

\bibitem{Deist2022midcktNA}
\bibinfo{author}{Deist, E.} \emph{et~al.}
\newblock \bibinfo{title}{Mid-circuit cavity measurement in a neutral atom array}.
\newblock \emph{\bibinfo{journal}{Phys. Rev. Lett.}} \textbf{\bibinfo{volume}{129}}, \bibinfo{pages}{203602} (\bibinfo{year}{2022}).

\bibitem{fowler2012surface}
\bibinfo{author}{Fowler, A.~G.}, \bibinfo{author}{Mariantoni, M.}, \bibinfo{author}{Martinis, J.~M.} \& \bibinfo{author}{Cleland, A.~N.}
\newblock \bibinfo{title}{Surface codes: Towards practical large-scale quantum computation}.
\newblock \emph{\bibinfo{journal}{Phys. Rev. A}} \textbf{\bibinfo{volume}{86}}, \bibinfo{pages}{032324} (\bibinfo{year}{2012}).

\bibitem{delfosse2021almost}
\bibinfo{author}{Delfosse, N.} \& \bibinfo{author}{Nickerson, N.~H.}
\newblock \bibinfo{title}{Almost-linear time decoding algorithm for topological codes}.
\newblock \emph{\bibinfo{journal}{{Quantum}}} \textbf{\bibinfo{volume}{5}}, \bibinfo{pages}{595} (\bibinfo{year}{2021}).

\bibitem{Kschischang2001productsum}
\bibinfo{author}{Kschischang, F.}, \bibinfo{author}{Frey, B.} \& \bibinfo{author}{Loeliger, H.-A.}
\newblock \bibinfo{title}{Factor graphs and the sum-product algorithm}.
\newblock \emph{\bibinfo{journal}{IEEE Transactions on Information Theory}} \textbf{\bibinfo{volume}{47}}, \bibinfo{pages}{498--519} (\bibinfo{year}{2001}).

\bibitem{noorshams2013productsum}
\bibinfo{author}{Noorshams, N.} \& \bibinfo{author}{Wainwright, M.~J.}
\newblock \bibinfo{title}{Belief propagation for continuous state spaces: Stochastic message-passing with quantitative guarantees}.
\newblock \emph{\bibinfo{journal}{Journal of Machine Learning Research}} \textbf{\bibinfo{volume}{14}}, \bibinfo{pages}{2799--2835} (\bibinfo{year}{2013}).

\bibitem{Grospellier2021combininghardsoft}
\bibinfo{author}{Grospellier, A.}, \bibinfo{author}{Grou{\`{e}}s, L.}, \bibinfo{author}{Krishna, A.} \& \bibinfo{author}{Leverrier, A.}
\newblock \bibinfo{title}{Combining hard and soft decoders for hypergraph product codes}.
\newblock \emph{\bibinfo{journal}{{Quantum}}} \textbf{\bibinfo{volume}{5}}, \bibinfo{pages}{432} (\bibinfo{year}{2021}).

\end{thebibliography}

\end{document}